\documentclass[11pt,a4paper]{iopart}
\usepackage{graphicx,color}
\usepackage{dcolumn}
\usepackage{epsfig}
\usepackage{setstack}

\usepackage{iopams}
\usepackage{euscript}
\usepackage{amssymb}
\usepackage{amsthm}
\usepackage{newlfont}
\usepackage{mathrsfs}
\usepackage{subfigure}
\usepackage{todonotes,cite}
\usepackage{subfigure}

\newcommand{\opunit}{\textrm{1}\kern-0.22em\textrm{l}}

\newcommand{\eg}{\textit{e.g.}}
\newcommand{\ie}{\textit{i.e.}}

\def\bea{\begin{eqnarray}}
\def\eea{\end{eqnarray}}
\def\ba{\begin{array}}
\def\ea{\end{array}}
\def\n{\nonumber}

\def\la{\langle}
\def\ra{\rangle}

\begin{document}
\title{Run-and-Tumble particles in Two-dimensions under Stochastic Resetting}
\author{Ion Santra$^1$, Urna Basu$^{1,2}$, Sanjib Sabhapandit$^1$}
%\author{Ion Santra}
%\ead{ion@rri.res.in}
%\author{Urna Basu}
%\author{Sanjib Sabhapandit}
\address{$^1$ Raman Research Institute, Bengaluru 560080, India}
\address{$^2$ S. N. Bose National Centre for Basic Sciences, Kolkata 700106, India}
\ead{ion@rri.res.in}

 \begin{abstract}
We study the effect of stochastic resetting on a run and tumble particle (RTP) in two spatial dimensions. We consider a resetting protocol which affects both the position and orientation of the RTP: with a constant rate the particle undergoes a positional resetting to a fixed point in space and orientation randomization. We compute the radial and $x$-marginal stationary state distributions and show that while the former approaches a constant value as $r \to 0$, the latter diverges logarithmically as $x \to 0.$ On the other hand, both the marginal distributions decay exponentially with the same exponent far away from the origin. We also study the temporal relaxation of the RTP and show that the position distribution undergoes a dynamical transition to a stationary state. We also study the first passage properties of the RTP in the presence of the resetting
and show that the optimization of the resetting rate can minimize the mean first passage time. We also give a brief discussion on the stationary states for resetting to the initial position with fixed orientation.
\end{abstract}

%\maketitle

\section{Introduction}
Active matter is a collection of   self-propelled or `active' agents, each of which can generate dissipative, persistent motion by extracting energy from their surroundings \cite{Romanzuk,Sriram,Marchetti2017,roadmapactive,Bechinger}. Dynamics of such systems are inherently non-equilibrium and lead to many remarkable features  which are strikingly different from their equilibrium counterparts.  Recent years have seen a tremendous surge of research work on collective and individual properties of active particles, which include flocking \cite{flocking1, flocking2}, phase separation \cite{separation1, separation2, separation3}, absence of well defined pressure  \cite{rtppressure} as well as non-Boltzmann stationary state and unusual relaxation behaviour \cite{nonBoltzman1,nonBoltzman2,nonBoltzman3,nonBoltzman4,nonBoltzman5,
nonBoltzman6,abp2018,rtp1d}.

Microscopically, active particle systems are  modeled by stochastic processes with correlated noise. One of the simplest model of an active particle is the so called Run and Tumble particle (RTP) which describes the overdamped motion of a particle  along an internal direction which itself changes stochastically \cite{berg,activrev}. Initially  introduced as a model for bacterial motion, RTP has become one of the most well studied active particle systems and shows a range of interesting statistical properties which include clustering at the boundaries of confining region\cite{Bechinger}, crossover from short-time ballistic to long-time diffusive behavior \cite{ion1}, non-Boltzmann steady state distributions in presence of traps \cite{rtp1dtrap,rtp2008,rtp_trap2,rtp_trap3}, universal behavior of marginal survival probability \cite{rtpddim}.

  Stochastic resetting refers to intermittent interruption and restart of a dynamical process \cite{resetting_review}. The paradigmatic example is that of a Brownian diffusive particle which is reset to its initial position with a certain rate\cite{res_diffusion,diff_arbitd}. This simple process leads to a set of interesting behaviors including a non-equilibrium stationary state, dynamical transition in the relaxation \cite{relaxation_ss} to it as well as a non-monotonic behaviour of mean first passage time \cite{searchoptimise,diff_optimalres,optimalreset1,fp_restart}.  Over the past decade, the effect of resetting has been studied in many variations and generalizations of simple diffusion. Specific examples include underdamped diffusion \cite{underdampeddiff}, Levy processes\cite{transitionlevyflight,optimal_firstarrival,reset_flightscales}, fractional Brownian motions \cite{fractionalbm} and random acceleration process \cite{rapreset}. Moreover, effect of various different protocols like non-Poissonian waiting time between consecutive resets \cite{intermittentreset,powerlawwait}, resetting with time-dependent rate \cite{timedepreset}, resetting in a confinement \cite{interval,circle} or to an extended region \cite{CCRW}, presence of a refractory period  \cite{refractorydiff} have also been studied in the context of diffusion or diffusion-like processes.

 A natural question is what is the effect of resetting when the underlying dynamics is active, instead of ordinary diffusion. This question has recently been studied in the context of active Brownian  particles \cite{scacchi,abpreset} and RTP in one dimension (1d) \cite{1drtpreset}. In this article, we study the effect of stochastic resetting on an RTP  in two spatial dimensions (2$d$). Unlike ordinary diffusion, for RTPs in higher dimensions, the different components of the position are not independent  and thus one would naturally expect a richer behavior.

In this work we present an analytical understanding of the effect of the resetting dynamics on the RTP. Evaluation of the moments indicate that in the presence of the stochastic resetting the RTP reaches a stationary state at large times. We study the approach to the non-equilibrium stationary state (NESS) and show that it undergoes a dynamical phase transition from a transient to a stationary state. We characterize the stationary state by computing the exact NESS position probability distributions. The radial distribution approaches a constant value dependent on the resetting rate as $r\to 0$, while the $x$-marginal distribution diverges logarithmically as $|x|\to 0$; both the marginal distributions decay exponentially with the same decay length at distances far away from the origin. We also look at the first passage properties by calculating the marginal survival probability of the RTP. Numerically we show the mean first passage time (MFPT) is minimized at some optimal resetting rate.

 The paper is organized as follows: the model is discussed in Sec.~\ref{model}; relevant results from earlier works are reviewed in Sec.~\ref{review}. The resetting protocol considered mainly in this article and the corresponding renewal equations are introduced in Sec.~\ref{ren1}; the stationary state distributions are computed in Sec.~\ref{sec:stationary}, while how the relaxation to the corresponding stationary state occurs is discussed in Sec.~\ref{transition}. First passage properties are investigated in Sec.~\ref{surv}. Finally we discuss some other possible resetting protocols in \ref{ren2} and conclude in Sec.~\ref{concl}.
  
\section{Model}\label{model}

We consider an overdamped run and tumble particle moving on the $x-y$ plane. The particle moves or `runs' with a constant speed $v_0$ along some internal orientation characterized by an angle $\theta$ and then `tumbles' to a new orientation, uniformly distributed in $[0,2\pi]$, and again runs along the new orientation with the speed $v_0$. The tumbling events occur at a constant rate $\gamma,$ \ie, the waiting time between two consecutive tumblings follows an exponential distribution. The Langevin equations describing the time evolution of the position of the particle are 
 \bea
 \dot{x}&=&v_0\cos\theta (t),\nonumber\\
 \dot{y}&=&v_0\sin\theta (t),
 \label{langevinwithoutreset}
 \eea
 where $\theta$ changes stochastically as described above. Equation~\eref{langevinwithoutreset} resembles a 2$d$ Brownian particle with effective noises 
  $\sigma_x(t)= \cos \theta(t)$ and $\sigma_y(t)= \sin \theta(t)$. However unlike a passive Brownian particle, the auto-correlation of this effective noise has an exponential form,
 \bea
 \langle \sigma_x(t)\sigma_x(0)\rangle=\langle \sigma_y(t)\sigma_y(0)\rangle=\frac{1}{2}e^{-\gamma t}.
 \eea
In this paper we add a stochastic resetting to this RTP dynamics: With rate $\alpha$ the particle  restarts the process starting from the same initial conditions. In the following we mostly consider the case where the new orientation $\theta'$ is chosen uniformly from $[0,2\pi]$ upon resetting. In Sec.~\ref{ren2} we consider the scenario where the orientation is reset to a fixed $\theta_r.$

\section{2$d$ RTP without resetting dynamics}\label{review}
In this section we recall the dynamics of the 2$d$ RTP in the absence of resetting and quote some relevant results from Refs. \cite{ion1,stadje,rtpddim} which we will use in the rest of the article. 

Let us consider an RTP starting from the origin, oriented along a random direction $\theta \in [0,2\pi],$ at time $t=0.$ The particle position evolves according to the Langevin equation \eref{langevinwithoutreset} --- at each tumbling event, the orientation $\theta$ changes to a new value $\theta',$ chosen from a uniform distribution in $[0,2\pi].$ Consequently, the position distribution remains isotropic at all times. The tumbling dynamics gives rise to an intrinsic time-scale  $\gamma^{-1}$ for the RTP which separates a short-time active ballistic and a long-time diffusive regime.
This crossover is visible from the variance of the position
 \bea
 \langle x^2(t)\rangle _0&=&\frac{v_0^2}{\gamma}\left ( t-\frac{1-e^{-\gamma t}}{\gamma}\right), \label{eq:xvarworesetting}
 \eea
which reduces to 
\bea
\langle x^2(t)\rangle _0&=& \left\{
\begin{array}{cc}
(v_0t)^2 & \text{for} ~~t \ll \gamma^{-1}, \cr
\frac{v_0^2}{\gamma} t & \text{for} ~~t\gg \gamma^{-1}.
\end{array}
\right.
\eea
Because of the isotropic nature of the motion, all the odd moments of the position components vanish. \\

\noindent {\bf Position distribution:} The Fokker Planck equation governing the time evolution of the position distribution of the RTP is given by
\bea
\fl \qquad  \frac{\partial}{\partial t} \cal{P}_0 (\vec{r},\theta,|\theta _0) &=&-\hat{n}.\vec{\nabla}\cal{P}_0(\vec{r},\theta,t |\theta_0) -\gamma \cal{P}_0(\vec{r},\theta,t|\theta_0)+ \gamma\int\frac{d\theta'}{2\pi}\cal{P}_0(\vec{r},\theta',t|\theta_0)
\eea
where $\cal{P}_0(\vec{r},\theta,t|\theta_0)$ denotes the probability that the particle is at $\vec{r},$ with orientation $\theta$ at time $t$, starting from origin and an initial orientation $\theta_0$ at $t=0;$ $\hat n= (\cos \theta, \sin \theta)$ is the unit vector along $\theta.$ The above equation can be solved exactly to obtain the radial probability distribution $\cal P_0(r,t)$ (as in \cite{ion1,stadje}),
\bea 
 \cal{P}_0 (r,t) &=&e^{-\gamma t}\bigg[\delta (r-v_0 t)+ \frac{\gamma r}{ v_0}\frac{\exp\bigg(\frac{\gamma}{v_0} \sqrt{v_0^2 t^2-r^2}\bigg)}{\sqrt{v_0^2 t^2-r^2}}\Theta (v_0 t-r) \bigg].~
 \label{rdistworeset}
 \eea
 where $\Theta(z)$ denotes the Heaviside function.  Note that this radial distribution  is normalized as $\int_0^\infty \cal P_0(r,t)dr=1.$  The $x$-marginal distribution can also be computed explicitly (as in \cite{ion1}),
\bea
\fl  P_0 (x,t) &=&e^{-\gamma t}\left( \frac{1}{\pi \sqrt{v_0^2t^2-x^2}}+\frac{\gamma}{2v_0} \Bigg[ L_0 \left(\frac{\gamma}{v_0} \sqrt{v_0^2t^2-x^2}\right) +I_0 \left(\frac{\gamma}{v_0} \sqrt{v_0^2t^2-x^2}\right)  \Bigg] \right),
\label{xdistworeset}
 \eea
where  $ {I_0(z)}$ is the the modified Bessel function of the first kind and $ {L_0(z)}$ is the modified Struve function \cite{dlmf}. \\ 
\noindent{\bf Survival probability:}  The survival probability $S_0(x_0,t)$ of an RTP in $d$-dimensions in the presence of an absorbing boundary at $x=0$ denotes the probability that the $x$-component of the displacement has not crossed the $x=0$ plane during the time interval $[0,t]$, starting from a given $x_0\geq 0.$ In particular for $x_0=0$, it can be computed explicitly \cite{rtp1d,rtpddim} and is given by
  \bea
  S_0(0,t)&=&\frac{e^{-\gamma t/2}}{2}\left[I_0\left(\frac{\gamma t}2\right)+I_1\left(\frac{\gamma t}2\right)\right].
  \label{survworeset}
  \eea
\section{Renewal Equation and Moments}\label{ren1}

In this section we study the motion of the 2$d$ RTP in the presence of a stochastic resetting. As introduced in Sec.~\ref{model}, the resetting is implemented by restarting the particle from origin with a constant rate $\alpha.$ Here we focus on the case where, after each resetting, the orientation is randomized and chosen uniformly from $[0,2\pi].$ Note that, in this  case, the effect of resetting on the orientation is same as that of a tumbling event.  

For this resetting mechanism, it is straightforward  to write a renewal equation for the position probability distribution,
 \bea
 \cal{P}_{\alpha}(\vec{r},t)&=&e^{-\alpha t}\cal{P}_0(\vec{r},t)+\alpha\int_{0}^{t} ds~e^{-\alpha s}\cal{P}_0(\vec{r},s).
\label{renewr}
 \eea
 where $\cal{P}_{\alpha}(\vec{r},t)$ and $\cal{P}_0(\vec{r},t)$ denote the position distributions in the presence and absence of resetting, respectively. The first term in the above equation corresponds to the situation where there are no resetting events during $[0,t].$  The second term contains the contribution from all the trajectories where the last resetting occurred at a time $t-s.$

In this work, we are particularly interested in the radial and $x$-marginal distributions in the presence of the resetting. It is easy to see that the marginal distributions also follow renewal equations  of the same structure. For example, for $\alpha >0,$ the marginal $x$ distribution evolves according to,
  \bea
 P_{\alpha}(x,t)&=&e^{-\alpha t}P_0(x,t)+\alpha\int_{0}^{t} ds~e^{-\alpha s}P_0(x,s).
\label{renewx}
 \eea
where $P_0(x,t),$ the distribution in the absence of resetting, is given by  \eref{xdistworeset}.

The $n^{th}$ moment of the position in the presence of resetting can be immediately computed by multiplying both sides of  \eref{renewx} by $x^n$ and integrating over $x$. Thus the general renewal equation for the moments is
 \bea
 \langle x^n(t)\rangle _{\alpha}&=&e^{-\alpha t}\langle x^n(t)\rangle _0 +\alpha\int_{0}^{t} ds~e^{-\alpha s}\langle x^n(s)\rangle_0
\label{momentsrenew} 
 \eea
 where $\la \cdot \ra_\alpha$ and $\la \cdot \ra_0$ denote statistical averages in the presence and absence of resetting, respectively.  The system is still isotropic and the odd moments vanish at all times. Let us look at the first non-zero moment, \ie,  the variance. Using $\langle x^2(s)\rangle$ from  \eref{eq:xvarworesetting}  in  \eref{momentsrenew} we have
 \bea
 \langle x^2(t)\rangle _{\alpha}
&=& \frac{v_0^2}{\alpha(\alpha+\gamma)} - \frac{v_0^2 e^{-\alpha t}}{\alpha\gamma(\alpha+\gamma)} \left(\alpha e^{-\gamma t}+ \alpha +\gamma \right).
 \eea
 Thus, at short times $(t\ll(\alpha+\gamma)^{-1})$ we have a ballistic behavior, $\langle x^2(t)\rangle _{\alpha}\approx v_0^2 t^2/2$, which is same as in the case without resetting. However, at large times $(t \gg \alpha^{-1})$ we see that the variance becomes time-independent, $\langle x^2(t)\rangle _{\alpha}\rightarrow \frac{v_0^2}{\alpha(\alpha+\gamma)}$ as $t\to\infty$. This indicates that the position distribution approaches a non-equilibrium stationary state at large times. In the following section we compute the marginal position distributions in the stationary state. 
 
\section{Stationary state distributions}
\label{sec:stationary}
In the presence of the stochastic resetting, the position distribution of the 2$d$ RTP evolves following the renewal equation \eref{renewr}.  As mentioned in the previous section, at late times, the distribution approaches a stationary limit, which is expected to be isotropic. In this section we investigate the nature of the stationary distributions for the radial and $x$-components of the position. 

\subsection{Radial Distribution}

\begin{figure}[h]
 \includegraphics[width=8 cm]{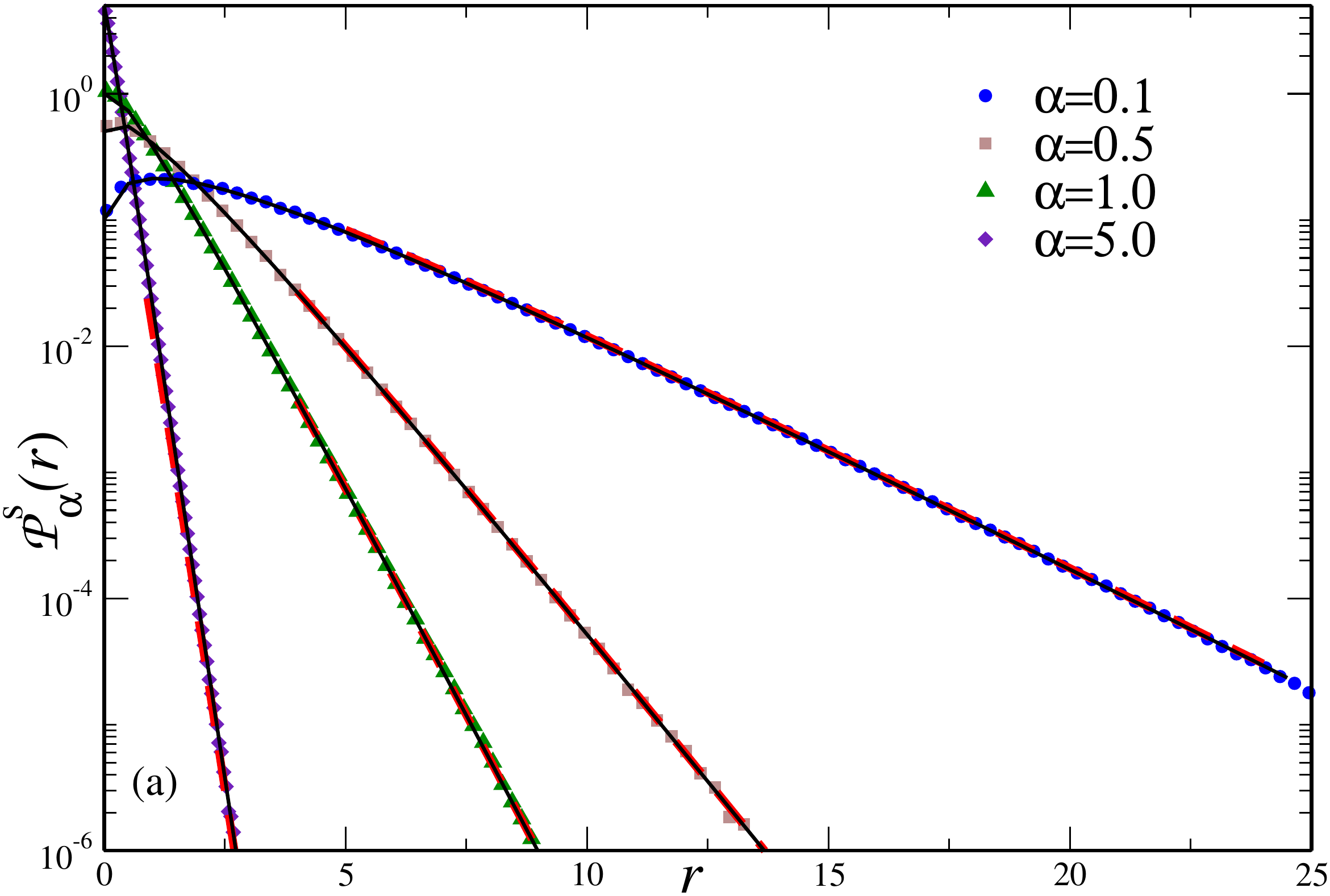}
 \includegraphics[width=8 cm]{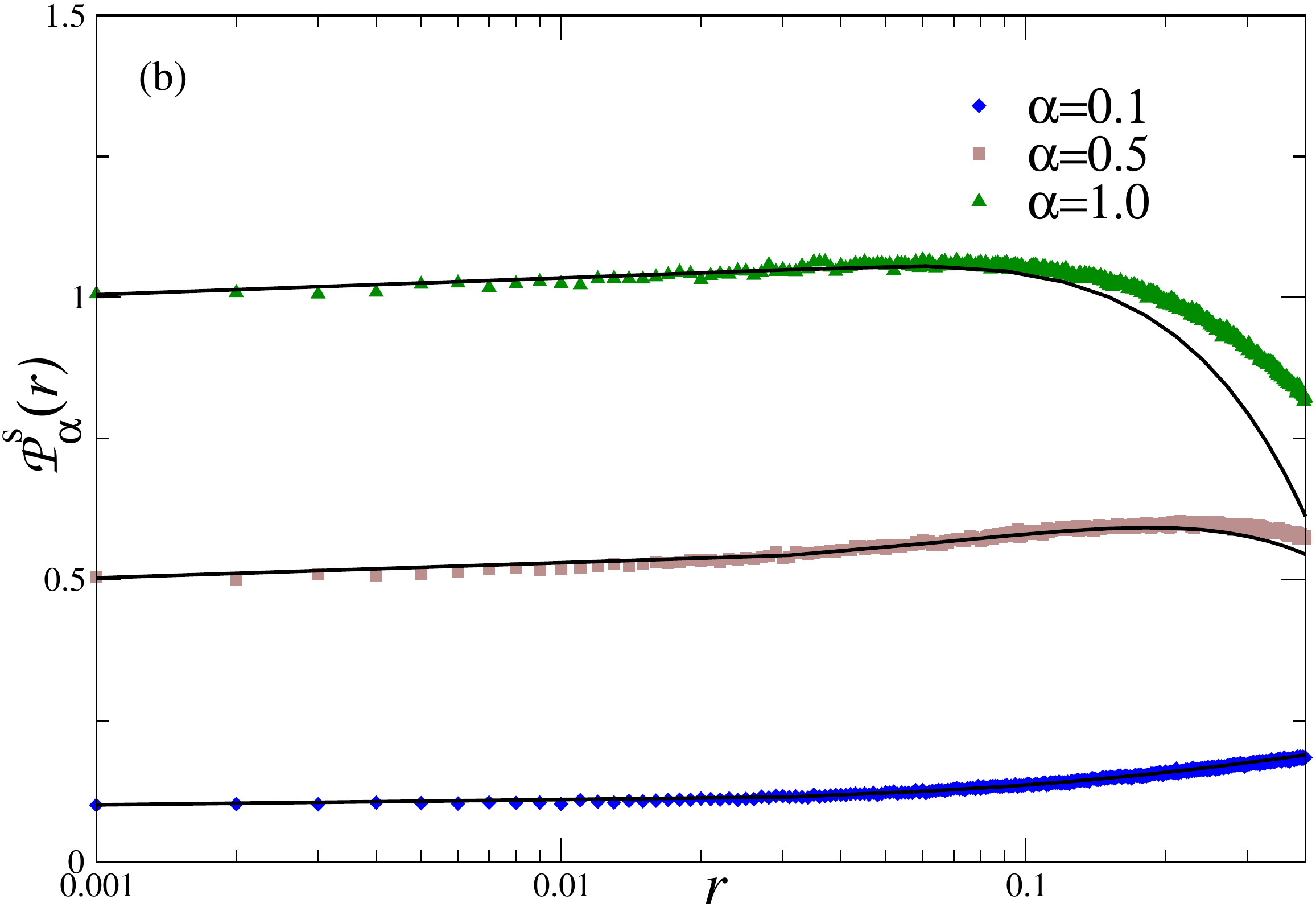}
\caption{Plot of stationary state radial distribution  $P^{\text{s}}_{\alpha}(r)$ as a function of $r$ for  different values of $\alpha$ and a fixed $\gamma=1.$ The symbols in both panels correspond to the data obtained from numerical simulations. In (a), the solid black lines are from the numerical evaluation of the series sum \eref{psrseries} keeping upto $n=60$ terms while the red dashed lines represent the large deviation function given by  \eref{rldf}. (b) shows the same distribution zoomed in near $r=0$ along with the theoretical prediction  \eref{pssrorigin} (solid black lines).}
\label{rstationary}
\end{figure}

Let us first look at the radial distribution $\cal P_\alpha(r,t) = \int_0^{2 \pi} d \phi ~r \cal P_\alpha(\vec r,t)$ where $\phi$ denotes the polar angle.
In the presence of the resetting, $\cal P_\alpha(r,t) $
follows a renewal equation similar to  \eref{renewr},
\bea
 \cal{P}_{\alpha}(r,t)&=&e^{-\alpha t}\cal{P}_0(r,t)+\alpha\int_{0}^{t} ds~e^{-\alpha s}\cal{P}_0(r,s).
\label{renewradial}
 \eea
 The stationary distribution is obtained by taking the $t\rightarrow\infty$ limit in the above equation,
 \bea
 \fl \cal{P}^{s}_{\alpha}(r)=\int_{0}^{\infty}ds~e^{-\alpha s}\cal{P}_0(r,s)\cr
 \fl \qquad \quad = \frac{\alpha}{v_0}e^{-(\alpha+\gamma) \frac r{v_0}}+\frac{\alpha\gamma r}{v_0}\int_{r/v_0}^{\infty} ds ~ e^{-(\alpha + \gamma)s} \frac{\exp\left(\frac \gamma {v_0} \sqrt{v_0^2 s^2- r^2}\right)}{\sqrt{v_0^2 s^2- r^2}}, 
\label{renewr2} 
 \eea
 where we have used \eref{rdistworeset} to arrive at the last line.
Let us denote the integral in the second term by $H(r).$ Using two successive variable transforms, $y = v_0s/r$ and $\omega = y-1,$ it reduces to,
 \bea
H(r)=\frac 1 {v_0}e^{-(\alpha+\gamma)\frac r{v_0}}\int_{0}^{\infty}d\omega ~
e^{-(\alpha + \gamma)\frac{r \omega}{v_0}} ~\frac{\exp{\left(-\frac{r\gamma}{v_0}\sqrt{\omega(\omega+2)}\right)}}{\sqrt{\omega(\omega+2)}}.
\label{prscaled}
 \eea
 This integral can be computed exactly by using the series expansion of the second exponential in the integrand, and integrating each term separately thereafter. This exercise leads to an exact expression for $H(r)$ as a sum of an infinite series,
\bea
H(r) = \frac{\sqrt \pi}{v_0} \sum_{n=0}^\infty  \frac {B_n}{n!} \left(\frac {2\gamma^2 r}{(\alpha+\gamma)v_0}~ \right)^{\frac n2} K_{\frac n 2} \Bigg(\frac{(\alpha +\gamma)r}{v_0} \Bigg). 
\eea 
where, $K_\nu(z)$ is the modified Bessel function of the second kind (see \cite{dlmf}) and,
\bea
B_n = \frac{\sec \frac{n \pi}2}{ \Gamma(\frac {1-n}2)} = \left \{ 
\begin{array}{ll}
\frac 1 \pi (\frac{n-1}2)! & \textrm{for odd} ~~ n \\[0.25 em]
\frac 1 {\sqrt\pi }\frac {n!}{2^n (n/2)!} & \textrm{for even} ~~ n.
\end{array} \label{eq:Bn}
\right. 
\eea
Finally, we have the stationary radial distribution,
\bea
\fl \qquad \cal{P}^{s}_{\alpha}(r)=\frac{\alpha}{v_0}e^{-(\alpha+\gamma)\frac{r}{v_0}}+\frac{\alpha\gamma \sqrt{\pi}r}{v_0^2}\sum_{n=0}^\infty  \frac {B_n}{n!} \left(\frac {2 \gamma^2 r}{(\alpha+\gamma)v_0}~ \right)^{\frac n2} K_{\frac n 2} \Bigg(\frac{(\alpha +\gamma)r}{v_0} \Bigg) ,
\label{psrseries}
\eea
where $B_n$ is given by  \eref{eq:Bn}. Figure~ \ref{rstationary} (a) compares this prediction with the data obtained from numerical simulations for $\cal{P}^{s}_{\alpha}(r)$  for different values of $\alpha$ for a fixed $\gamma;$ the solid lines correspond to the analytical prediction  \eref{psrseries}, with the sum truncated after a few terms, and the  symbols correspond to the numerical simulation results. This Figure~ illustrates that the series converges pretty fast, and can be used to compute  stationary distribution at any $r$  to arbitrary accuracy.

It is interesting to look at the asymptotic behavior of $\cal{P}^{s}_{\alpha}(r).$
Using the series expansion of $K_{\frac n 2}(z)$ near $z=0,$ we get, for small $r,$ 
\bea
\fl \qquad \cal{P}^{s}_{\alpha}(r)=\frac{\alpha}{v_0}-\frac{\alpha\gamma}{v_0^2}r \ln r +\frac{\alpha\gamma r}{v_0^2} \left( \gamma\ln\frac{\alpha}{2v_0}+\gamma(\Gamma_{\text E}+1) +\alpha \right)+\cal{O}(r^2 \ln r),
\label{pssrorigin}
\eea
where 
$\Gamma_{\text E}$ is the Euler-Mascheroni constant. This is compared with the numerical simulations in  Figure~\ref{rstationary}(b).

Next we look at the large $r$ behavior of the stationary state distribution. It is difficult to extract the large $r$ behavior directly from  \eref{psrseries}; instead we  recast  \eref{prscaled} in a different form,
 \bea
H(r)=\frac{1}{v_0}e^{-(\alpha+\gamma)\frac r {v_0}}\int_{0}^{\infty}d\omega\frac{e^{-\frac{r}{v_0}\Lambda(\omega)}}{\sqrt{\omega(\omega+2)}}, \n
 \eea
where $\Lambda(\omega)=(\alpha+\gamma)\omega-\gamma\sqrt{\omega(\omega+2)}.$ 
It is straightforward to check that $\Lambda(\omega)$ is a non-monotonic function of $\omega$ with a minimum at $\omega_0= \frac{\alpha +\gamma}{\sqrt{\alpha^2 +2\alpha \gamma}}-1$. Thus, for large $r,$ the above integral 
can be evaluated using saddle point method (See \ref{saddlept}), which yields
\bea
H(r) \approx \left(\frac{2\pi }{v_0 r\sqrt{\alpha^2+2\alpha\gamma}}\right)^{1/2} \exp{\left[-\frac r {v_0}~ \sqrt{\alpha^2+2\alpha\gamma}\right]}.
\eea
 Since this exponential decays much slower than the first term in  \eref{renewr2}, the large $r$ behavior of the radial distribution is dominated by this term, and we have,
\bea
\cal{P}^{s}_{\alpha}(r) \approx \frac{\alpha\gamma}{v_0^{3/2}}\left(\frac{2\pi r}{\sqrt{\alpha^2+2\alpha\gamma}}\right)^{1/2} \exp{\left[-\frac r {v_0}~ \sqrt{\alpha^2+2\alpha\gamma}\right]}.
\label{rldf}
\eea 
This exponential decay for large $r$ is compared with the results from numerical simulations in  Figure~\ref{rstationary}(a) which shows an excellent agreement.

\subsection{Marginal $x$--distribution}
In the absence of resetting, the $x$-marginal distribution of the 2$d$ RTP shows an   algebraic divergence near the boundaries $|x|=v_0 t.$  It is interesting to see how  this marginal distribution is affected by the introduction of stochastic resetting. 
To calculate the stationary $x$-marginal distribution, we take the $t\rightarrow\infty$ limit of \eref{renewx},
\bea
\fl P^{s}_{\alpha}(x)=\alpha\int_0^{\infty}ds~e^{-\alpha s}P_0(x,s)=\alpha\int_{|x|/v_0}^{\infty} ds~\frac{e^{-(\gamma+\alpha)s}}{\pi\sqrt{v_0^2s^2-x^2}}\cr+\frac{\alpha\gamma}{2v_0}\int_{|x|/v_0} ^\infty ds~e^{-(\gamma +\alpha) s}\left[L_0\left(\frac{\gamma}{v_0}\sqrt{v_0^2 s^2-x^2}\right)+I_0\left(\frac{\gamma}{v_0}\sqrt{v_0^2 s^2-x^2}\right)\right].
\eea
The integral in the first term can be computed exactly and yields $\frac{1}{\pi v_0}K_0\left(\frac{(\alpha+\gamma) |x|}{v_0} \right)$. Let the integral in the second term be denoted by $G(x)$. Thus,
\bea
P^{s}_{\alpha}(x)=\frac{\alpha}{\pi v_0}K_0\left(\frac{(\alpha+\gamma) |x|}{v_0} \right)+\frac{\alpha\gamma}{2v_0}G(x).
\label{pssxf}
\eea 
Using two successive transformations $y=v_0 s/|x|$ and $\omega=y-1$, $G(x)$ reduces to
\bea
\fl G(x)=\frac{|x|}{v_0}e^{-(\alpha+\gamma)\frac{|x|}{v_0}}\int_{0}^{\infty} d\omega~ e^{-(\alpha+\gamma)\frac{|x|\omega}{v_0}}\left[L_0\left(\frac{\gamma |x|}{v_0}\sqrt{\omega(\omega +2)}\right)+I_0\left(\frac{\gamma |x|}{v_0}\sqrt{\omega(\omega +2)}\right)\right].~\cr
\label{pssxf2}
\eea

\begin{figure}[h]
\includegraphics[width=8 cm]{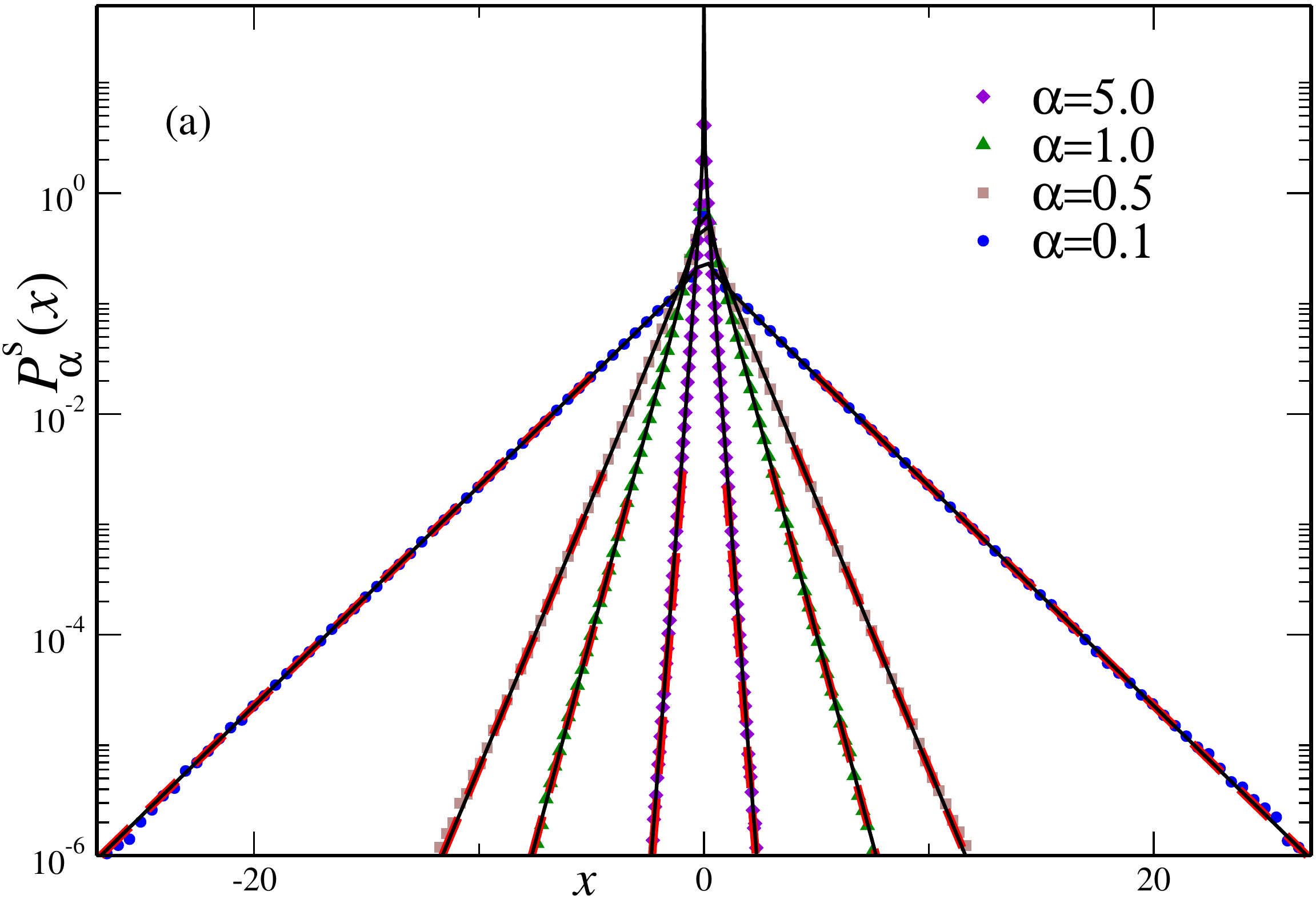}
\includegraphics[width=8 cm]{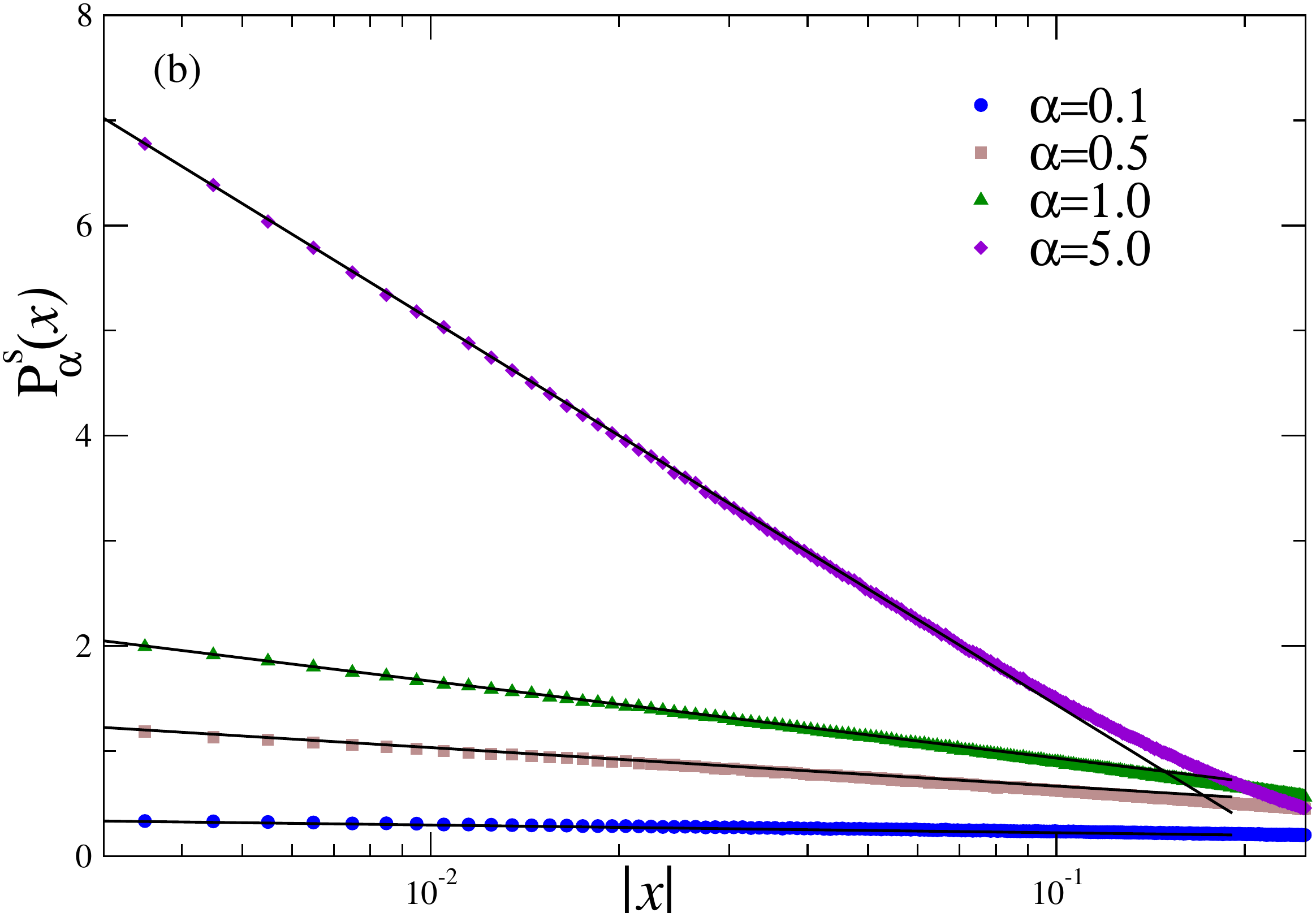}
\caption{Plot showing  stationary state marginal $x$ distribution for $\gamma=1$ and different values of $\alpha$. The colored points in both Figure~s correspond to the data obtained from numerical simulation. In (a), the solid black lines are from the numerical evaluation of the series Eq~\ref{psxseries} for $n=60$ while the red dashed lines represent the large deviation functions as given by  \eref{xldf}. Sub Figure~ (b) compares the behavior of the stationary state distribution as given by  \eref{pssxorigin} (indicated by solid black lines) to the results obtained from numerical simulation.}
\label{xstationary}
\end{figure}

This integral can be performed if we use the series expansions of Struve and Bessel functions (sections 11.2.2 and 10.25.2 of \cite{dlmf}) and integrate each term separately. This leads to an infinite series form for $G(x)$,
\bea
\fl G(x)=\frac{|x|}{\sqrt{\pi}v_0}\sum_{n=0}^{\infty}\left(\frac{\gamma^2 |x|}{2 v_0 (\alpha+\gamma)}\right)^{n}\left[\frac{1}{n!} \sqrt{\frac{2 v_0}{(\alpha+\gamma)|x|}} ~ K_{n+1/2}\left(\frac{(\alpha+\gamma)|x|}{v_0}\right)\right.\cr
\fl \qquad \qquad \qquad \qquad \quad \quad + \left.\frac{\gamma}{(\alpha+\gamma)} \frac 1{\Gamma(n+3/2)} ~K_{n+1}\left(\frac{(\alpha+\gamma)|x|}{v_0}\right)\right].
\label{psxseries}
\eea
The complete $x$-marginal distribution is then given by  \eref{pssxf} along with  \eref{psxseries}; in fact, the series sum converges fast and can be used to compute the stationary state marginal $x$-distribution up to any arbitrary accuracy. Figure~ \ref{xstationary}(a) compares the predicted stationary distribution with the data obtained from numerical simulations.

For small $x$, the leading order behavior of $P_\alpha^s(x)$ can be obtained if we use the asymptotic expressions of $K_{\nu}(z)$ for small $z$. Doing this exercise, we see that $G(x)$ approaches an $x$-independent finite value as $x\to 0.$ On the other hand, the first term in \eref{pssxf}  diverges in this limit, as $K_0(z) \sim - \log z$  as $z \to 0.$ Combining, we get, for $|x|\to 0,$
\bea
P^{s}_{\alpha}(x) = -\frac{\alpha}{\pi}\ln |x| +  \cal{O}(1). \label{pssxorigin}
\eea
Thus the stationary state distribution has a logarithmic divergence near the origin; see  Figure~\ref{xstationary}(b).
Next we turn our attention to the tails of the stationary state distribution. Extracting the large $x$ behavior from  \eref{psxseries} is very cumbersome, so we again use a saddle point method. For large $|x|$ we can use the asymptotic forms of the Struve and Bessel functions in Eq~\eref{pssxf2}. Thus we have,
\bea
 G(x) &=&2\sqrt{\frac{|x|}{v_0\gamma}} e^{-(\alpha+\gamma)\frac{|x|}{v_0}} \int_{0}^{\infty} d\omega~ e^{-(\alpha+\gamma)\frac{|x|\omega}{v_0}}\frac{\exp \left[\frac{\gamma |x|}{v_0}\sqrt{\omega(\omega+2)}\right]}{\sqrt{2\pi\sqrt{\omega(\omega+2)}}}\cr
 &=&2\sqrt{\frac{|x|}{v_0\gamma}} e^{-(\alpha+\gamma)\frac{|x|}{v_0}}\int_{0}^{\infty} d\omega~ \frac{e^{-\frac{|x|}{v_0}\Lambda(\omega)}}{\sqrt{2\pi\sqrt{\omega(\omega+2)}}}, \label{eq:H_saddle}
\eea
where $\Lambda(\omega)=(\alpha+\gamma)\omega-\gamma\sqrt{\omega(\omega+2)}$, which clearly has a minimum w.r.t. $\omega$. Thus at large $|x|$, we can again use the saddle point method to evaluate the integral in \eref{eq:H_saddle},
\bea
G(x) \approx\frac{2}{\sqrt{\alpha^2+2\alpha\gamma}}\e^{-\frac{|x|}{v_0}\sqrt{\alpha^2+2\alpha\gamma}}.
\eea
Putting this back in  \eref{pssxf}, and using the asymptotic form $K_0(z)\sim z^{-1/2}e^{-z}$ for large $z$, we see that the decay length in the first term is always much smaller than that in $G(x)$. Thus at large $|x|$, the stationary state distribution decays as,
\bea
P^{s}_{\alpha}(x)\approx \frac{\alpha\gamma}{v_0\sqrt{\alpha^2+2\alpha\gamma}}\e^{-\frac{|x|}{v_0}\sqrt{\alpha^2+2\alpha\gamma}}.
\label{xldf}
\eea
This behavior  is compared with numerical simulations in  Figure~\ref{xstationary}(a)(red dashed lines).

 It is interesting to compare the stationary state distributions obtained here with those studied previously in the context of diffusion \cite{res_diffusion,diff_arbitd}.  For a passive diffusive particle in $d=2$, it has been shown that the presence of stochastic resetting results into a radial distribution which vanishes at the origin $\sim r\ln r$ \cite{diff_arbitd}. On the other hand, the marginal $x$-distribution, in that case, attains a finite value at the origin, while showing an exponential decay for all $|x|>0$ \cite{res_diffusion}. In contrast, here we see that, the introduction of stochastic resetting to an underlying 2$d$ RTP dynamics leads to a finite value of the radial probability density at $r=0$ (see   \eref{pssrorigin}). On the other hand, 
the $x$-marginal distribution for the RTP shows a logarithmic divergence near $x=0$ in the presence of resetting (see \eref{pssxorigin}). In short, we see that the behavior of the stationary state distributions of the RTP near the origin is significantly different   than its passive counterpart.

Physically, this difference can be understood  form the following argument. For both the radial and $x$-marginal distributions,  the leading contribution near the origin comes from the trajectories which undergo none or very few tumblings between two consecutive resetting events. For example, for the radial distribution, 
the non-zero contribution at the origin  comes from the first term of  \eref{rdistworeset} which is actually the position distribution of the free RTP in the short time active regime. Similarly, for the $x$-marginal distribution, the divergence near the origin arises from the first term in \ref{xdistworeset} which again, corresponds to the trajectories which undergo resetting with none or very few tumblings. Such trajectories with small number of tumblings carry the signature of the active nature of the underlying system, which, in turn shows up in the stationary state distribution in the presence of resetting.

On the other hand, the tails of both the marginal distributions decay exponentially with the same exponent $\sqrt{\alpha^2+2 \alpha \gamma}.$ In fact, this exponent is identical to the one obtained in \cite{1drtpreset} in the context of resetting of 1d RTP, and thus appears to be robust in any dimensions.

It is useful to consider some special limiting scenarios.

\begin{itemize}

\item{\it Diffusive limit:}  In the absence of resetting, the RTP dynamics reduces to ordinary diffusion in the limit $\gamma\to \infty,$ $v_0\to\infty$ but with a finite ratio $\frac{v_0^2}{2\gamma}=D_{\textrm{eff}}$ which plays the role of an effective diffusion constant. It is easy to see that, in this limit, both the radial and $x$-marginal distributions for the RTP reduce to the corresponding known results for diffusive particles. For example, using the limit $\gamma\to \infty,$ $v_0\to\infty,$ and finite $D_{\textrm{eff}}$ in \eref{renewr2} we have,
\bea
 \fl \qquad \cal{P}^{s}_{\alpha}(r)=\frac{\alpha r}{2D_{\textrm{eff}}}\int_{0}^{\infty} \frac{ds}s ~ e^{-\alpha s} \exp{\left(-\frac{r^2}{4D_{\textrm{eff}}~ s}\right)} = \frac{\alpha r}{D_{\textrm{eff}}}K_0\left(\sqrt{\frac{\alpha}{D_{\textrm{eff}}}}r\right),
 \eea
which is identical to the result obtained in \cite{diff_arbitd}. Similarly, the $x$-marginal distribution \eref{pssxf} reduces to a pure exponential in the diffusive limit, which coincides with the well known result obtained in \cite{res_diffusion}. 

\item{\it Small flip rate $\gamma \to 0$:}
In this limit the second term in the expression for both radial and $x$-marginal distributions \eref{renewr2} and \eref{pssxf} goes to zero. Thus we find that the stationary state distributions decay exponentially with a decay constant $\frac{v_0}{\alpha}$ which is the mean distance traveled by the particle between two consecutive resetting events when there are no flips. 

\end{itemize}
\section{Relaxation to stationary state and position distributions}
\label{transition}
 It is interesting to look at how the non-equilibrium stationary state as described in the previous section is attained. In this section we look at how the radial and marginal position distributions relax to the respective stationary state distributions.

 \subsection{Radial Distribution}
 We start from the renewal equation \eref{renewradial}, using  \eref{rdistworeset}, we have 
\bea
\fl \cal{P}_{\alpha}(r,t)=e^{-(\alpha +\gamma)t}\delta (r-v_0t)+\frac{\gamma r e^{-t(\alpha+\gamma-\gamma \sqrt{1- (r/v_0 t)^2})}}{v_0^2 t\sqrt{1- (r/v_0 t)^2}}+\frac{\alpha}{v_0 }e^{-(\alpha+\gamma)r/v_0}+ H(r,t)
\label{renrfull}
 \eea 
where, 
\bea
H(r,t) = \frac{\alpha\gamma r}{v_0^2}\int_{\frac r {v_0}}^t ds \frac{e^{-s\left(\alpha+\gamma-\gamma \sqrt{1-(r/v_0 s)^2}\right)}}{s\sqrt{1-(r/v_0 s)^2}}. \label{eq:Hrt_defn}
\eea 
 To evaluate the integral in  $H(r,t),$ we make a change of variable $s=t\tau$ and obtain,
 \bea
 H(r=zv_0 t,t)&=&\frac{\alpha\gamma zv_0 t}{v_0^2}\int_z^{1} d\tau~e^{-(\alpha+\gamma) t\tau }\frac{\exp\left[\gamma t\sqrt{\tau^2-z^2}\right]}{\sqrt{\tau^2-z^2}}\nonumber\\
 &=&\frac{\alpha\gamma z v_0 t}{v_0^2}\int_{z}^{1} d\tau~e^{-t\phi(z,\tau) }\frac{1}{\sqrt{\tau^2-z^2}}
\label{radsaddle}
 \eea
where $z=r/v_0 t$, ($z \in [0,1]$) and $\phi(z,\tau)=(\alpha+\gamma)\tau-\gamma\sqrt{\tau^2-z^2}$. Now at very large $t$ and fixed $z$ we can estimate $H(r,t)$ by saddle point method (worked out in details in \ref{saddlept}). The dominant contribution to the integral comes from the minimum of $\phi(z,\tau)$ at $\tau_0=\frac{z(\alpha+\gamma)}{\sqrt{\alpha^2+2\alpha\gamma}}$. Now, there can be two possibilities:
\begin{itemize}
\item $\tau_0 <1$: In this case the minimum of $\phi(z,\tau)$ lies within the limits of integration $(z,1)$ (note that $\tau_0>z$). Thus for $\tau_0<1$, 
\bea
 H(r=zv_0 t,t)&\approx&\frac{\alpha \gamma\sqrt{zv_0t} \sqrt{\pi/2}}{v_0^2(\alpha^2+2\alpha\gamma)^{1/4}}\e^{-zt\sqrt{\alpha^2+2\alpha\gamma}}.
\label{lr}
\eea
We drop the prefactors going forward as we are only interested in the behavior at the tails. Thus from  \eref{renewradial} and  \eref{renrfull}, we have for the region $\tau_0<1$, \ie, for $z<\frac{\sqrt{\alpha^2+2\alpha\gamma}}{\alpha+\gamma}$,
\bea
\cal{P}_\alpha(z,t)\sim \e^{-zt\sqrt{\alpha^2+2\alpha\gamma}}.
\label{rp1}
\eea
Note that the third term on the rhs of  \eref{renrfull} has been dropped as the corresponding length scale $\frac{v_0}{\alpha+\gamma}$ is much smaller than that in $H(r,t)$ which is $v_0\sqrt{\alpha^2+2\alpha\gamma}/(\alpha+\gamma)$.
\item $\tau_0>1$: In this case, the minimum of $\phi(z,\tau)$ lies outside the limits of integration, the minimum value of $\phi(z,\tau)$ within the integration limits is at the boundary $\tau=1$.Thus the dominant contribution to the integral comes from near $\tau=1$,
 \bea
 H(r=zv_0 t,t) \sim e^{-(\alpha+\gamma)t+\gamma t\sqrt{1-z^2}}.
 \eea 
  This is of the same order as the second term on the rhs of   \eref{renrfull}, which indicate to the fact that this contribution physically corresponds to the trajectories that have undergone none or very few resettings until time $t$. 

\end{itemize}  
  
\begin{figure}[h]
\includegraphics[width=6 cm]{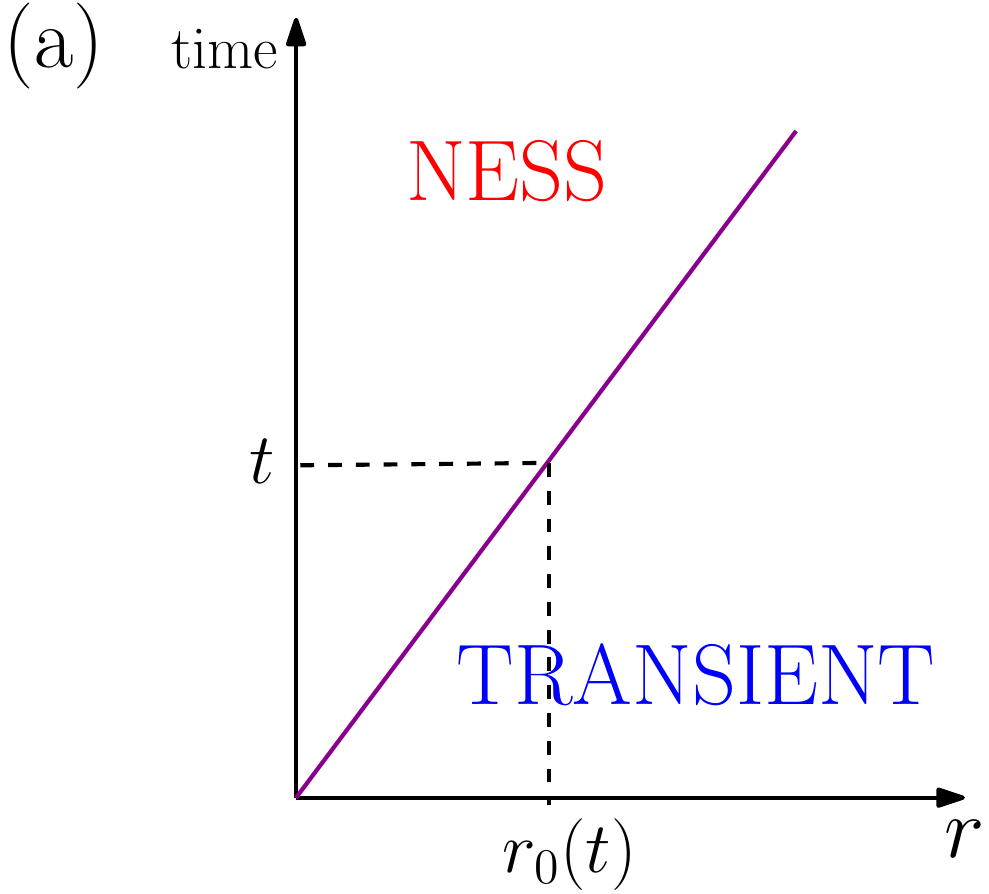}
~~~~~~\includegraphics[width=8 cm]{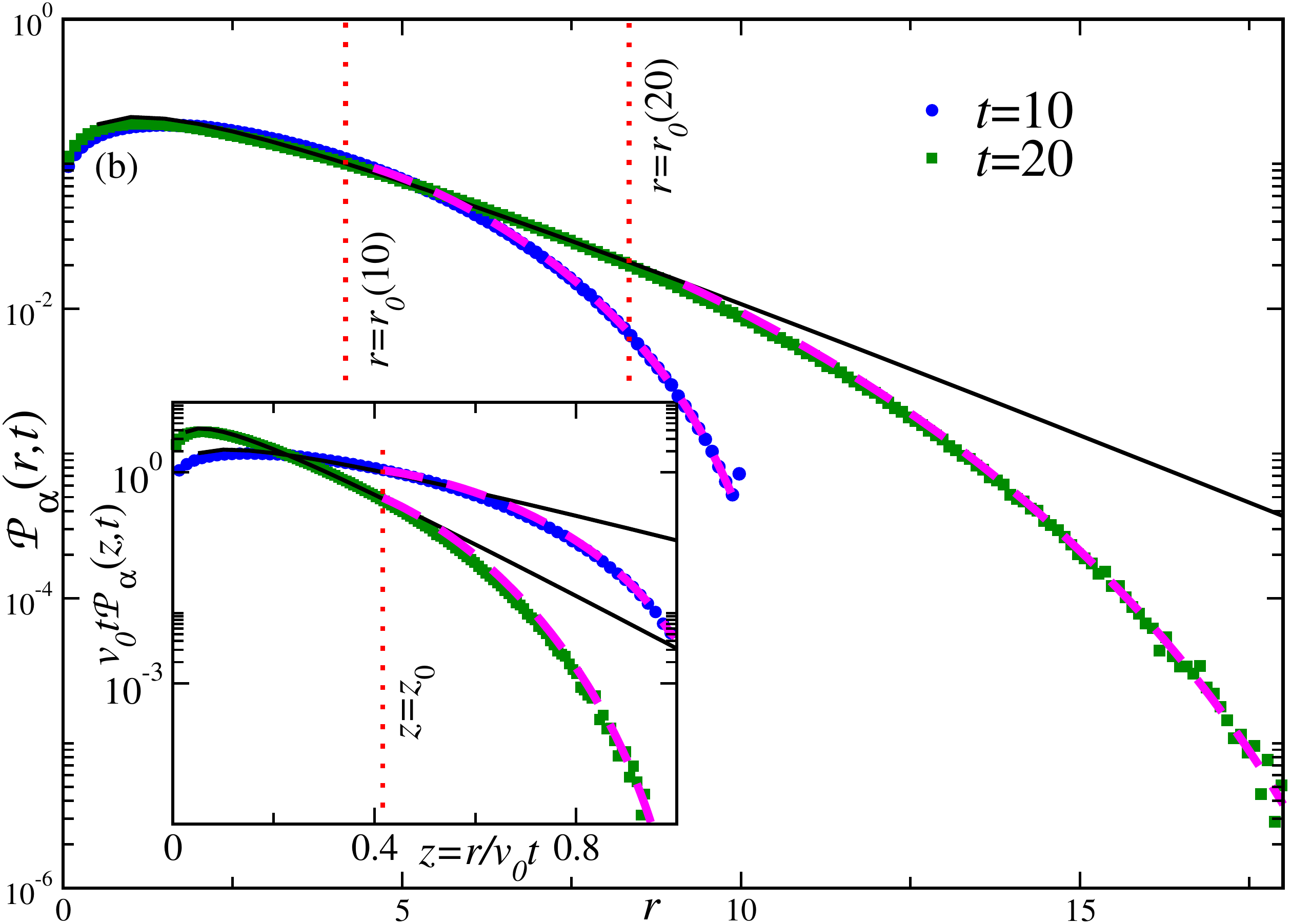}

\caption{Approach to the non-equilibrium stationary state of the radial distribution: (a) Diagrammatic representation of how the relaxation occurs in  the $r$-space; (b) Comparison of the numerical simulation results with our analytical predictions for $\gamma=1$ and $\alpha=0.1$. The symbols represent the data obtained from numerical simulation and the solid black lines indicate the large deviation form \eref{rldf}.  The dashed magenta lines represent the transient part of the distribution \eref{transitionz} and \eref{transitionr} with proper prefactors( see \ref{saddleapp1}). The main plot shows the radial distribution; the dashed red vertical lines indicate $r=r_0(t)$ for two values of $t=10,20$. The plot in the inset shows the distribution in terms of the scaled variable $z$, the dashed red vertical line denotes $z=\frac{\sqrt{\alpha^2+2\alpha\gamma}}{\alpha+\gamma}$.}
\label{nessclip}
\end{figure}

  So for the region $\tau_0>1$, \ie, $z>\frac{\sqrt{\alpha^2+2\alpha\gamma}}{\alpha+\gamma}$
  \bea
  \cal{P}_{\alpha}(z,t)\sim e^{-(\alpha+\gamma)t+\gamma t\sqrt{1-z^2}}.
   \label{rp2}  
  \eea
 Combining \eref{rp1} and \eref{rp2}, we have the distribution for the scaled variable 
\bea
 \cal{P}_{\alpha}(z,t)\sim \e^{-t \psi(z)}
\eea 
 where 
 \bea
 \psi(z)= \Bigg \{ 
\begin{array}{cc}
z\sqrt{\alpha^2+2\alpha\gamma} & ~ \textrm{for}~ z<\frac{\sqrt{\alpha^2+2\alpha\gamma}}{\alpha+\gamma},\cr
(\alpha+\gamma)-\gamma \sqrt{1-z^2} & ~ \textrm{for}~  z>\frac{\sqrt{\alpha^2+2\alpha\gamma}}{\alpha+\gamma}.
\end{array}
\label{transitionz}
  \eea 
Writing in terms of the original variable $r=z v_0 t$, this translates to,
  \bea
 \cal{P}_{\alpha}(r,t)\sim \Bigg \{ 
\begin{array}{cc}
e^{-\frac r {v_0}\sqrt{\alpha^2+2\alpha\gamma}} & ~ \textrm{for}~ r<r_0(t),\cr
e^{-(\alpha+\gamma)t+\gamma \sqrt{t^2-(r/v_0)^2}} & ~ \textrm{for}~  r>r_0(t).
\end{array}
\label{transitionr}
  \eea 
  where $r_0(t)=\frac{\sqrt{\alpha^2+2\alpha\gamma}}{\alpha+\gamma} t.$
Thus we see that at a large time $t$, the position distribution for the region $r<r_0(t)$ is time independent and has the exact same form as the stationary state large deviation function in  \eref{rldf}, while for the region $r>r_0(t)$ the distribution is time dependent.  
Since $r_0(t)$ is linear in $t$, the region which has reached stationary state grows at a constant speed as shown in  Figure~\ref{nessclip}(a). The relaxation of the position distribution calculated from the numerical simulations is compared with the our results \eref{transitionr} in  Figure~\ref{nessclip}(b). 

\subsection{Marginal $x$-distribution}

\begin{figure}[h]
\centering
\includegraphics[width=6 cm]{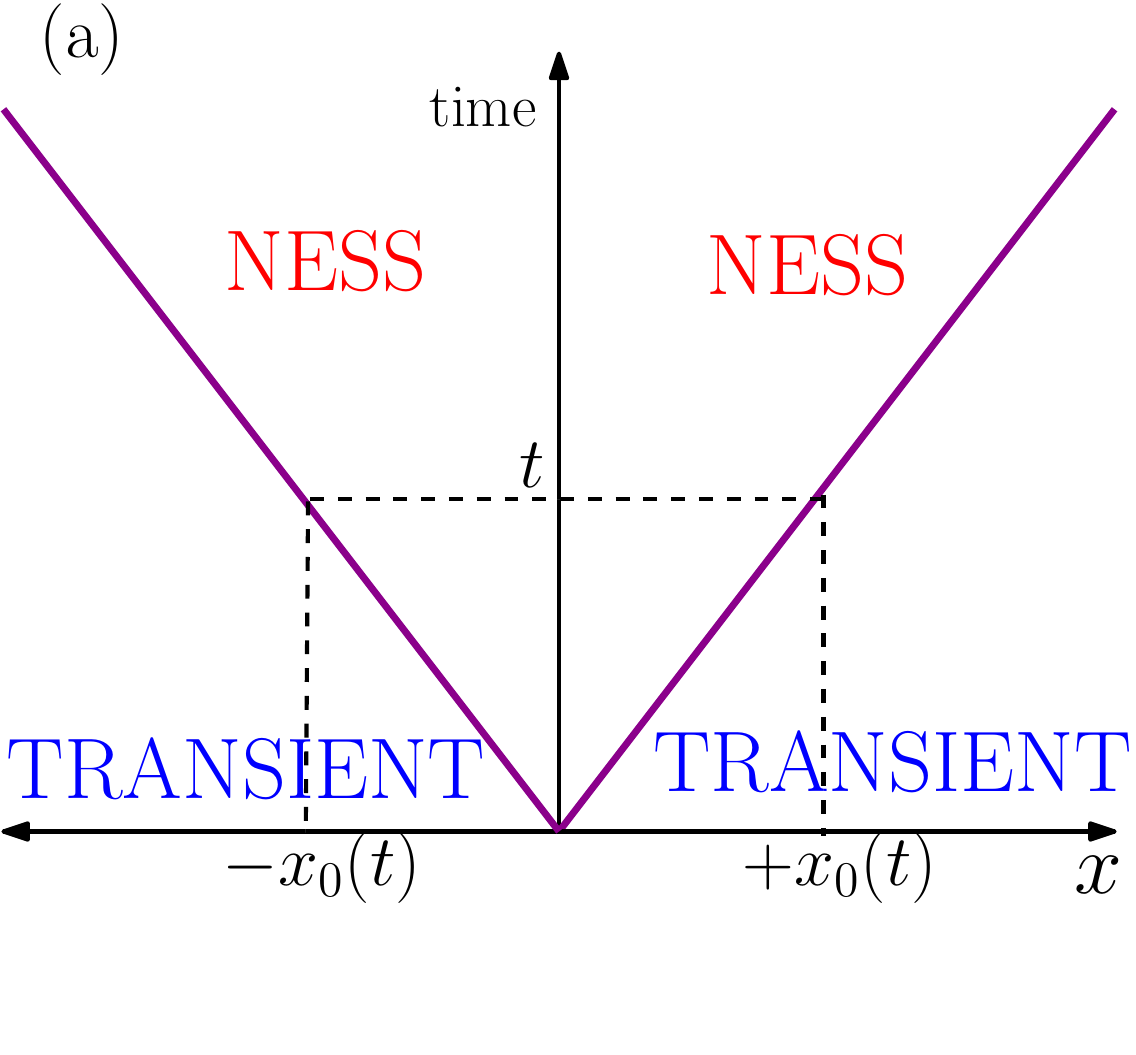}
~~~\includegraphics[width=8 cm]{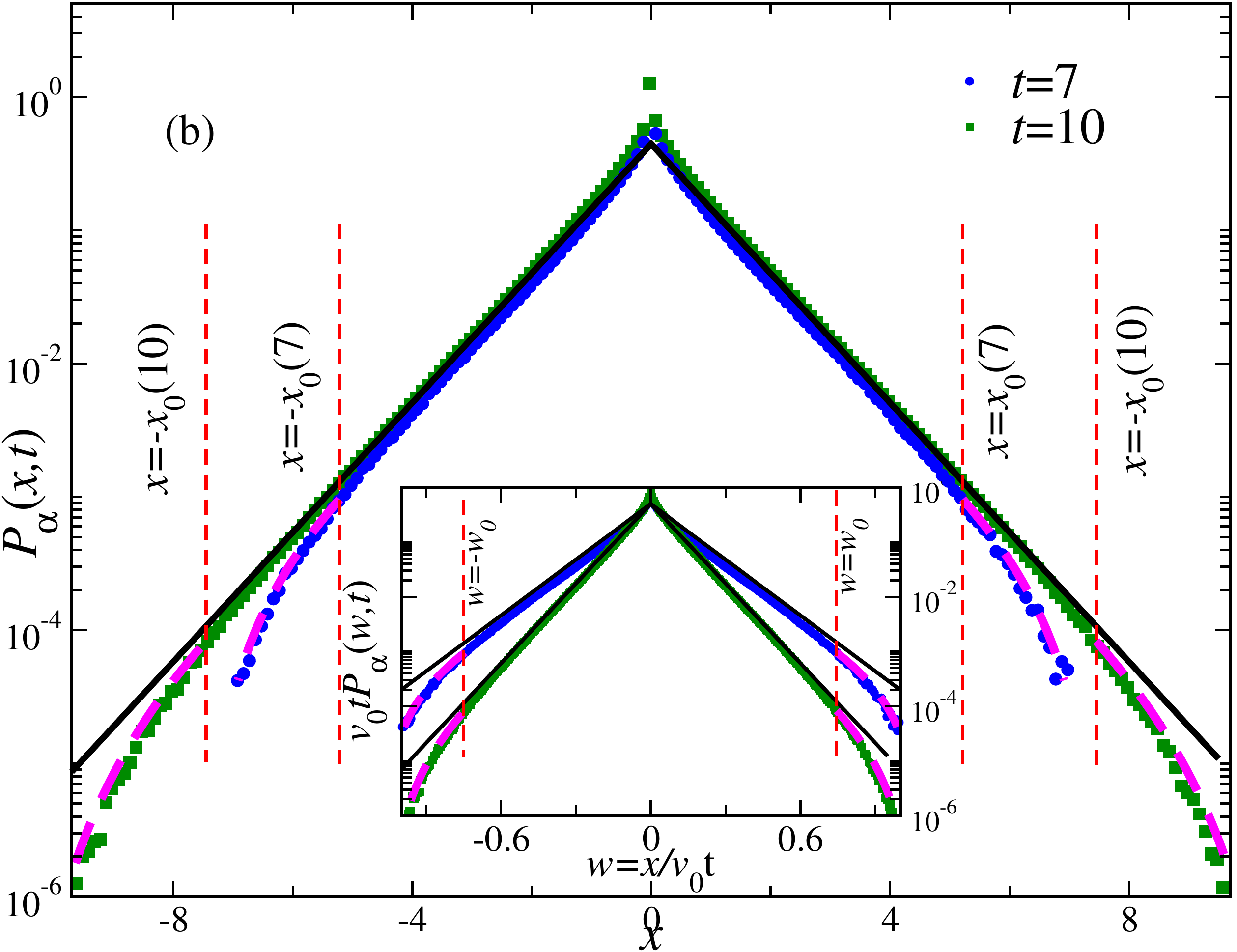}
\caption{Approach to the non-equilibrium stationary state of the $x$-marginal distribution: (a) Diagrammatic representation of how the relaxation occurs in  the $x$-space; (b) Comparison of the numerical simulation results with our analytical predictions with $\gamma=1$ and $\alpha=0.5$. The symbols represent the data obtained from numerical simulation and the solid black lines indicate the large deviation form \eref{xldf}. The dashed magenta lines represent the transient part of the distribution \eref{transitionw} and \eref{transitionxx} with proper prefactors. The main plot shows the radial distribution; the dashed red vertical lines indicate $r=r_0(t)$ for two values of $t=7, 10$. The plot in the inset shows the distribution in terms of the scaled variable $z$, the dashed red vertical line denotes $z=\frac{\sqrt{\alpha^2+2\alpha\gamma}}{\alpha+\gamma}$.}
\label{transitionx}
\end{figure}

It is also interesting to look at how the tails of the marginal $x$-distribution relax to  \eref{xldf}. We start from  \eref{renewx}; using the free propagator  \eref{xdistworeset}, we have,
\bea
\fl P_{\alpha}(x,t)=\frac{e^{-(\alpha+\gamma)t}}{\pi\sqrt{v_0^2 t^2-x^2}}+\frac{\gamma e^{-(\alpha+\gamma)t}}{2v_0}\left[ L_0 \left(\frac{\gamma}{v_0} \sqrt{v_0^2t^2-x^2}\right) +I_0 \left(\frac{\gamma}{v_0} \sqrt{v_0^2t^2-x^2}\right)  \right]\cr+\alpha\int_{\frac{|x|}{v_0}}^t ds~\frac{e^{-(\gamma+\alpha)s}}{\pi\sqrt{v_0^2s^2-x^2}}~+~G(x,t)
\label{xssexact}
\eea
where,
\bea
\fl G(x,t)=\frac{\alpha\gamma}{2v_0}\int_{\frac{|x|}{v_0}} ^t ds~e^{-(\gamma +\alpha) s}\left[L_0\left(\frac{\gamma}{v_0}\sqrt{v_0^2 s^2-x^2}\right)+I_0\left(\frac{\gamma}{v_0}\sqrt{v_0^2 s^2-x^2}\right)\right].
\eea
The integral in the third term on the rhs of  \eref{xssexact} can be done exactly and yields $\frac{\alpha}{v_0\pi}K_0\left(\frac{\gamma+\alpha}{v_0}|x|\right)$. To evaluate the integral in $G(x,t)$ we define a change of variable $s=t\tau$ and get,
\bea
\fl \qquad G(|x|=wv_0t,t)=\frac{\alpha\gamma t}{2v_0}\int_{\frac{|x|}{v_0}} ^1 d\tau~e^{-(\gamma +\alpha) t\tau}\left[L_0\left(\gamma t\sqrt{\tau^2-w^2}\right)+I_0\left(\gamma t\sqrt{\tau^2-w^2}\right)\right]\nonumber
\eea
For large $t$, we can use the asymptotic expressions for $L_0(z)$ and $I_0(z)$ for large $z$ \cite{dlmf}, to get
\bea
\fl G(|x|=wv_0t,t)\approx  \frac{\alpha \sqrt{\gamma t}}{v_0}\int_w^1 d\tau\frac{e^{-(\gamma+\alpha)t\tau+\gamma t\sqrt{\tau^2-w^2}}}{\sqrt{2\pi\sqrt{\tau^2-w^2}}}
=\frac{\alpha \sqrt{\gamma t}}{v_0}\int_w^1 d\tau\frac{e^{-t\Phi(w,\tau)}}{\sqrt{2\pi\sqrt{\tau^2-w^2}}},
\eea
where $w=|x|/v_0t$ and $\Phi(w,\tau)=(\gamma+\alpha)\tau-\gamma\sqrt{\tau^2-w^2}$. Now, at large $t$ and fixed $z$, we can estimate the value of the integral using a saddle point integral. The contribution to the integral comes from the minimum of $\Phi(w,t)$ w.r.t. $\tau$. The minimum $\tau_0=\frac{w(\alpha+\gamma)}{\sqrt{\alpha^2+2\alpha\gamma}}$. There can be two possibilities:
\begin{itemize}

\item $\tau_0<1$: In this case the minimum of $\Phi(w,t)$ lies within the integration limits $w,~1$. Thus we have,
\bea
G(|x|=wv_0t,t)\approx \e^{-wt\sqrt{\alpha^2+2\alpha\gamma}}.
\eea
Thus from  \eref{renewx} and  \eref{xssexact}, we have for the region defined by $\tau_0<1$,
\bea
P_{\alpha}(w,t)\sim \e^{-wt\sqrt{\alpha^2+2\alpha\gamma}}.
\label{xldf1}
\eea
Note that the third term in  \eref{xssexact}, $\sim K_0\left((\alpha+\gamma)\frac{|x|}{v_0}\right)$ has been dropped because the corresponding characteristic length scale, $\frac{v_0}{(\alpha+\gamma)}$, is much smaller than that in $G(x,t)$, $\frac{v_0\sqrt{\alpha^2+2\alpha\gamma}}{\alpha+\gamma}$. 
\item $\tau_0>1$: The minimum of $\Phi(w,t)$ lies outside the integration limits. The minimum value of $\Phi(z,t)$ within the integration limits $[z,1]$ is at the boundary $\tau=1$. Thus we have
\bea
G(|x|=wv_0t,t)\sim \e^{-(\gamma+\alpha)t+\gamma t\sqrt{1-w^2}}
\eea
 which is of the same order as the no-resetting terms in  \eref{xssexact}, thus indicating that this contribution comes from the trajectories which have under gone none or very few resettings.
  So for the region $\tau_0>1$, \ie, $w>\frac{\sqrt{\alpha^2+2\alpha\gamma}}{\alpha+\gamma}$
  \bea
  P_{\alpha}(w,t)\sim e^{-(\alpha+\gamma)t+\gamma t\sqrt{1-w^2}}.
   \label{xp2}  
  \eea
\end{itemize}
 Combining \eref{xldf1} and \eref{xp2}, we have
\bea
P_{\alpha}(w,t)\sim e^{-t\psi(w)}
\eea
where
\bea
\psi(w)= \Bigg \{ 
\begin{array}{cc}
w\sqrt{\alpha^2+2\alpha\gamma} & ~ \textrm{for}~ w<\frac{\sqrt{\alpha^2+2\alpha\gamma}}{\alpha+\gamma}\cr
(\gamma+\alpha)-\gamma \sqrt{1-w^2} & ~ \textrm{for}~  w>\frac{\sqrt{\alpha^2+2\alpha\gamma}}{\alpha+\gamma}.
\end{array}
\label{transitionw} 
\eea
Writing in terms of the original variables $x$ and $t$,
\bea
P_{\alpha}(x,t)\sim \Bigg \{ 
\begin{array}{cc}
e^{-\frac{|x|}{v_0}\sqrt{\alpha^2+2\alpha\gamma}} & ~ \textrm{for}~ |x|<x_0(t)\cr
e^{-(\alpha+\gamma)t+\gamma \sqrt{t^2-(x/v_0)^2}} & ~ \textrm{for}~  |x|>x_0(t).
\end{array}
\label{transitionxx} 
\eea
where $x_0(t)=\frac{\sqrt{\alpha^2+2\alpha\gamma}}{\alpha+\gamma}t.$ Thus, for the region $|x|<x_0(t)$, we find that, at large but finite $t$, the distribution is time independent and has the exact same form as the stationary state  \eref{xldf}, while for the region $|x|>x_0(t)$ the distribution is explicitly time dependent. This implies that the region $|x|<x_0(t)$ has relaxed to stationary state at time $t$ and since $x_0(t)$ is linear in $t$, the region which has reached stationary state grows at a constant speed as shown in  Figure~\ref{transitionx}(a). The relaxation of the position distribution as obtained in  \eref{transitionxx} is compared with the results of numerical simulations in  Figure~\ref{transitionx}(b).

\section{First Passage Properties}
\label{surv}
In this section we discuss the probability of survival of a RTP in presence of an absorbing boundary and the mean first passage time. The survival probability $S(x_0,t)$  of an RTP,  with an absorbing boundary at $x=x_{\text{abs}}$, is defined as the probability that starting from an initial position $x_0,$  the RTP has not crossed the absorbing boundary up to time $t$. In the context of a search process $x=x_{\text{abs}}$ is the target; and an event in which the RTP reaches the line $x=x_{\text{abs}}$ corresponds to the searcher successfully locating the target.

\begin{figure}[h]
\includegraphics[width=8 cm]{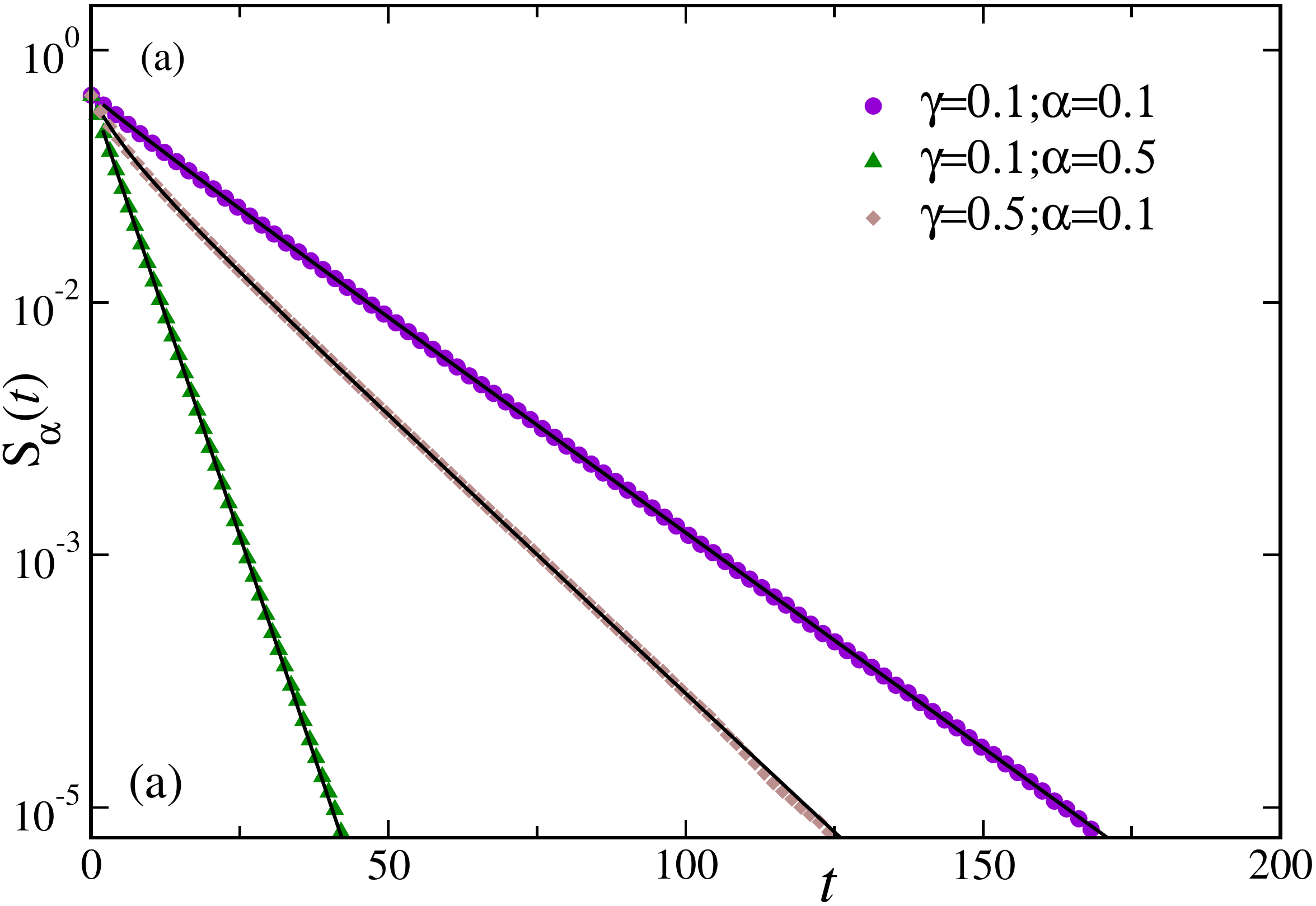}
\includegraphics[width=8 cm]{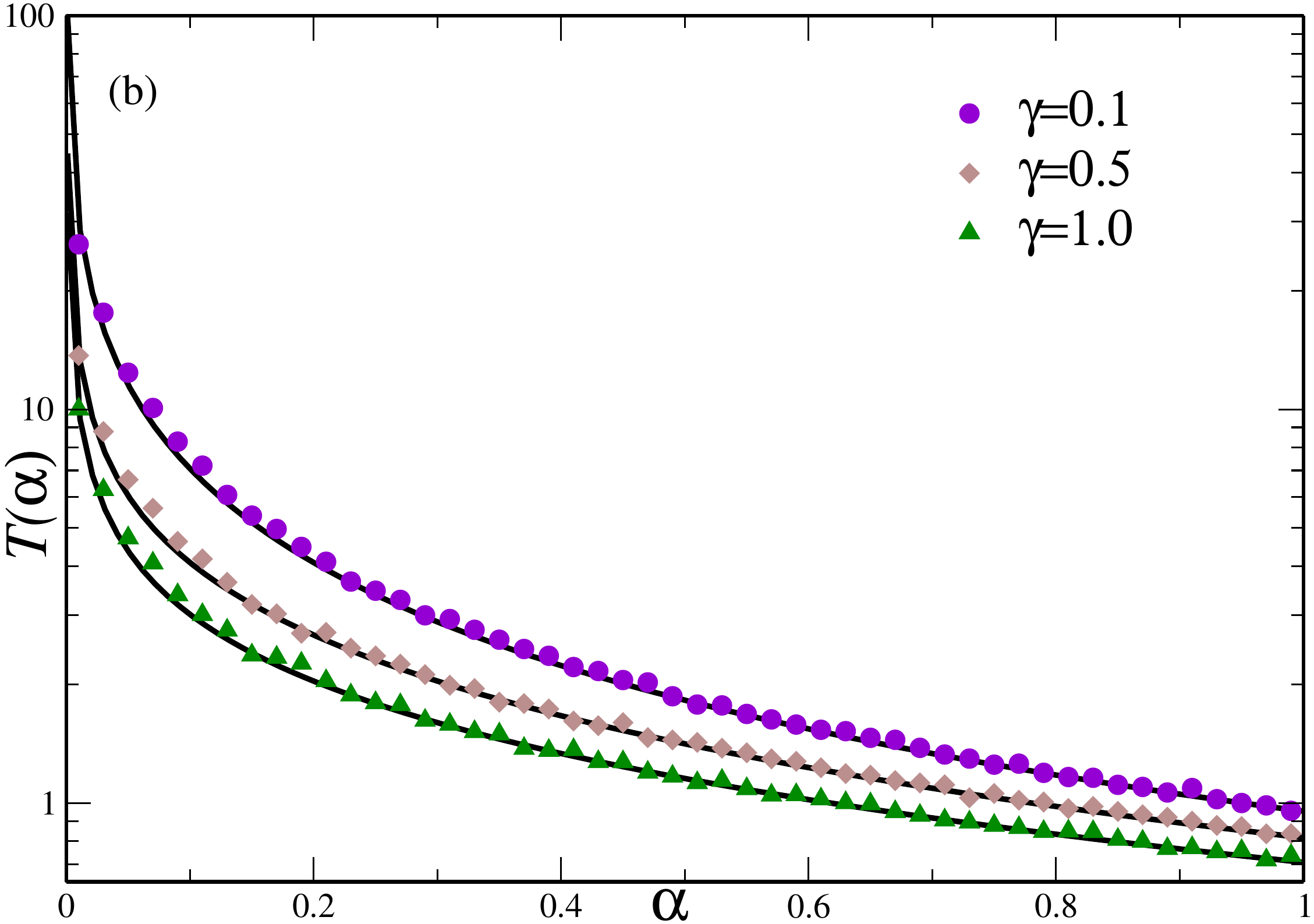}
\caption{Plot of first passage properties for $x_{\text{abs}}=x_r=0$. (a) Survival probability: Colored symbols denote results obtained from numerical simulation, while solid black lines correspond to  \eref{survreset}. (b) Mean first passage time: Colored points denote results obtained from numerical simulation, while solid black lines correspond to  \eref{mfptdef}.}
\label{survivalfig}
\end{figure}

In dimensions greater than one, the calculation of the survival probability of an RTP is difficult to calculate because the orientation is a continuous variable. However for the special case $x_0=x_{\text{abs}}=0$ the survival probability $S(0,t)$ of a d-dimensional RTP was calculated in \cite{rtpddim}. In the following we use the result obtained in \cite{rtpddim} (see  \eref{survworeset}) to investigate the survival probability for a 2$d$ RTP under resetting. 
 We begin by writing down a renewal equation for the survival probability $S_{\alpha}(x_0,t)$ for the protocol of resetting the position of the particle to the some point $(x_r,y_r)$ and randomizing the orientation (velocity) at each reset event.
\bea
S_{\alpha}(x_0,t)&=&e^{-\alpha t}S_{0}(x_0,t)+\alpha\int_{0}^{t}ds~e^{-\alpha s}S_0(x_r,s)S_{\alpha}(x_0,t-s).
\label{renewsurv1}
\eea
 where $S_0$ denotes the survival probability without resetting. The first term on the RHS is due to the trajectories which have not undergone resetting. The second term, on the other hand, integrates over all those survived trajectories where the last resetting occurred at a time $t-s,$  which accounts for the factor $\alpha\e^{-\alpha s}$. We consider $x_{\text{abs}}=x_r=0.$  Taking a Laplace transform, $\tilde{S}_{\alpha}(x,s)=\int_0^{\infty}dt~e^{-st} S_\alpha(x,t)$, on both sides of  \eref{renewsurv1} 
and setting the initial position $x_0=0,$ we get, 
\bea
\tilde{S}_{\alpha}(0,s)&=&\frac{\tilde{S}_0(0,\alpha+s)}{1-\alpha\tilde{S}_0(0,\alpha+s)}. \label{renewsurv2}
\eea
We are now in a position to use  \eref{survworeset}. Taking a Laplace transform $t\rightarrow s$, we have  
  \bea
  \tilde{S}_0(0,s)&=&\frac{1}{\sqrt{s(\gamma+s)}}.
  \eea
  Putting this is in  \eref{renewsurv2}, we get the survival probability in $s$-space as
  \bea
  \tilde{S}_{\alpha}(0,s)&=&\frac{1}{s+\sqrt{(s+\alpha)(s+\alpha+\gamma)}}.
  \label{survlaplace}
  \eea
   To invert the Laplace transform we need to write the corresponding Bromwich integral,
\bea
S_\alpha(0,t) = \frac 1{2 \pi i} \int_{c-i \infty}^{c+i \infty} ds~ \frac{e^{st}}{s+\sqrt{(s+\alpha)(s+\alpha+\gamma)}}
\eea   
where $c$ is chosen such that all the singularities of the integrand lie to the left of the line Re$[s]=c.$ Clearly, the above integral involves a branch-cut along the real $s$-axis in addition to a simple pole at $s=-\frac{\alpha(\gamma+\alpha)}{2\alpha+\gamma}.$ Taking into account all the contributions,  we finally have,
   \bea
   S_{\alpha}(0,t)&=&\frac{2\alpha(\gamma+\alpha)}{(\gamma+2\alpha)^2} e^{-\frac{\alpha(\gamma+\alpha)}{2\alpha+\gamma}t}+\frac{e^{-\alpha t}}{\pi}\int_{0}^{\gamma}du~ \frac{e^{-ut}\sqrt{u(\gamma-u)}}{\alpha^2+(2\alpha +\gamma)u}.
   \label{survreset}
   \eea
The second term involves a convergent integral, which unfortunately does not yield any closed form solution. If we write the numerator of the second term as an infinite series and do the $u$-integral we have the full survival probability as,
\bea
\fl S_{\alpha}(0,t)=\frac{2\alpha(\gamma+\alpha)}{(\gamma+2\alpha)^2} e^{-\frac{\alpha(\gamma+\alpha)}{2\alpha+\gamma}t}+\frac{\e^{-(\alpha+\gamma) t}\gamma^2}{2\alpha^2 \sqrt{\pi}}\sum_{n=0}^{\infty}\left(\frac{2\gamma(\alpha+\gamma)}{\alpha^2}\right)^n\Gamma\left[n+\frac 3 2\right] {}_1\tilde{F}_1\left[\frac 3 2,n+3,\gamma t\right] \cr
\label{survexact}
 \eea 
 where ${}_1\tilde{F}_1\left[a,b,z\right]$ is the regularized Kummer function \cite{dlmf}.  Using the asymptotic expansion of  ${}_1\tilde{F}_1\left[a,b,z\right]$ for small $z$ \cite{dlmf}, we have for small $t$,
\bea
S_{\alpha}(0,t)=\frac{1}{2}-\frac{1}{8} (2 \alpha +\gamma )t+\cal{O}(t^2).
\eea
  At $t=0$, the survival probability has the expected value $\frac 1 2,$  since we start with uniform initial conditions (\ie, the initial orientation is chosen uniformly from $[0,2\pi]$).
  Again for large $t$, we can use the asymptotic expansion of  ${}_1\tilde{F}_1\left[a,b,z\right]$ for large $z$ \cite{dlmf}. This yields, for large $t$,
  \bea
  S_\alpha(0,t)=\frac{2\alpha(\gamma+\alpha)}{(\gamma+2\alpha)^2} \e^{-\frac{\alpha(\gamma+\alpha)}{2\alpha+\gamma}t}+\cal{O}[\e^{-\alpha t}].
\label{survres4}  
  \eea
   Thus, at large times the survival probability decays as,
   \bea
  S_{\alpha}(0,t)\sim \e^{-\frac{\alpha(\gamma+\alpha)}{2\alpha+\gamma}t}.
\label{survres3}  
  \eea
 Equation \eref{survreset} is compared to numerical simulations in  Figure~~\ref{survivalfig}(a). They show excellent match. 

 A related observable is the first passage time, which is the time at which the particle reaches $x=x_{\text{abs}}$ for the first time. The corresponding first passage probability $F_{\alpha}(x_0,t)dt$ denotes the probability that the particle, starting at $x=x_0$ is absorbed at $x=x_{\text{abs}}$ during the time interval $[t,t+dt]$. It is the time derivative of survival probability,
 \bea
F_{\alpha}(x_0,t)&=&-\frac{\partial }{\partial t}S_{\alpha}(x_0,t).
\label{fptdef}
 \eea
  The mean first passage time (MFPT) is defined as the mean time taken to be absorbed and is thus given by,
  \bea
  T_{\alpha}(x_0)&=&\int_{0}^{\infty}dt~t F_{\alpha}(x_0,t)=\int_{0}^{\infty}dt~S_{\alpha}(x_0,t),
  \label{mfptdef}
  \eea
where, to obtain the second equation, we have used $F_{\alpha}(x_0,t)$ from  \eref{fptdef} and then performed an integration by parts. We also used the fact that $S_{\alpha}(x_0,\infty)\rightarrow 0$. The RHS of equation  \eref{mfptdef} is actually the Laplace transform $\tilde{S}_{\alpha}(x_0,s=0)$. For $x_0=0$ we can use  \eref{survlaplace} to get the MFPT. Thus we have,
\bea
T_{\alpha}(0)=\frac{1}{\sqrt{\alpha(\alpha+\gamma)}}.
\label{mfptexp}
\eea
This diverges as $\alpha\rightarrow 0$ and decreases monotonically with $\alpha$. This is due to the fact that every time we reset the particle back to the origin its orientation is chosen uniformly between $[0,2\pi]$, so the probability that it gets absorbed at a reset event is always half. The result obtained in \eref{mfptexp} is compared with numerical simulations in \ref{survivalfig}(b).

\begin{figure}
\includegraphics[width=8 cm]{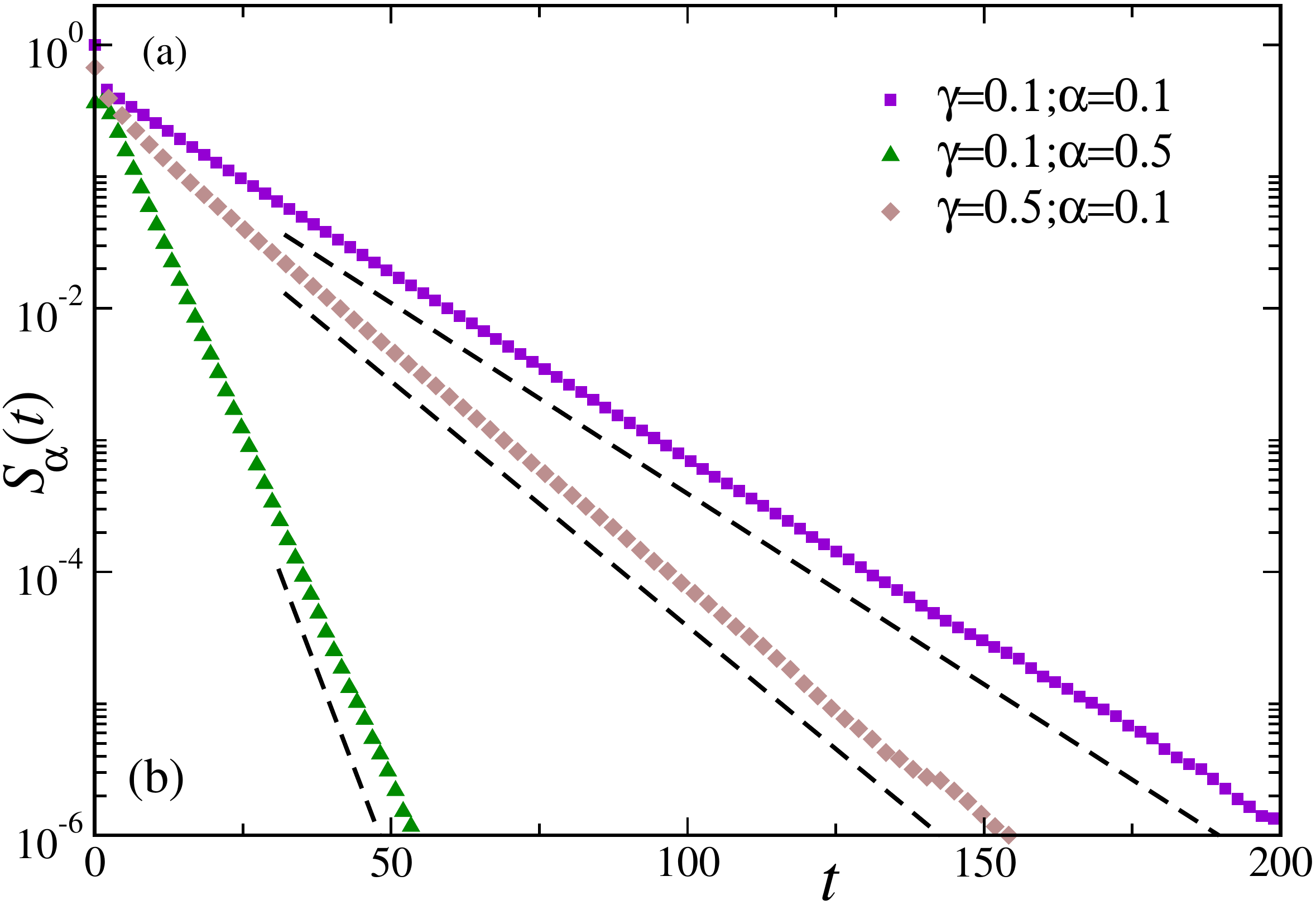}
\includegraphics[width=8 cm]{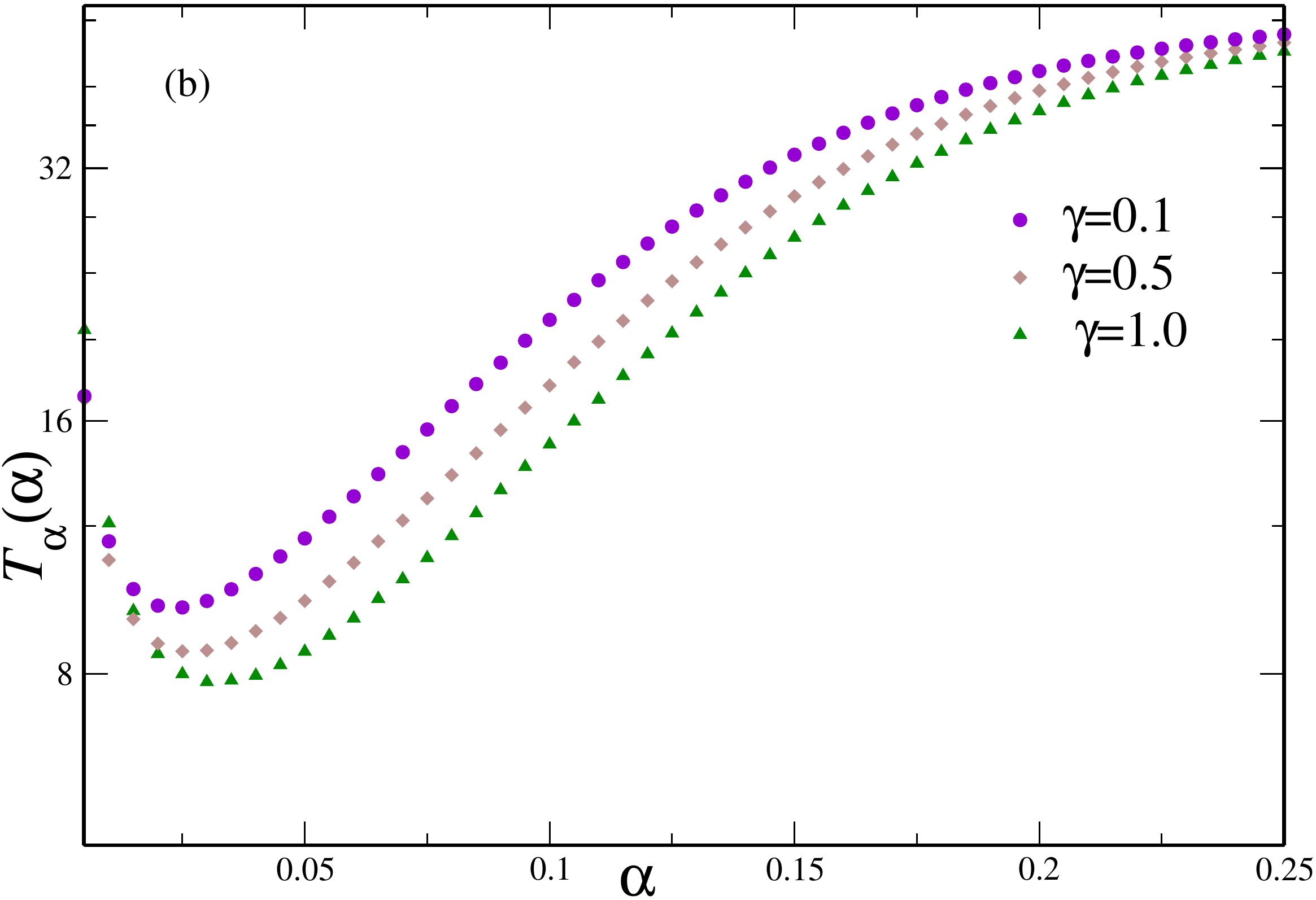}
\caption{Plot for $x_{\text{abs}}=-0.1$ and $x_r=0$: (a) Survival probability: Colored points denote results obtained from numerical simulation, black dashed line are the exponential tails as denoted by  \eref{survres3}. (b) Mean first passage time: Colored points are obtained from numerical simulation. The existence of a minimum in all the curves clearly indicate the fact that the mean first passage time is minimized for an optimal value of the resetting rate $\alpha$. }
\label{survivalfig2}
\end{figure}

It is interesting to see what happens when we push the absorbing boundary parallel to $y$-axis to some negative $x$ (\ie, $x_{\text{abs}}<x_r=0$). The problem with solving the backward FP equation does not allow us to analytically find how the survival probability will change in that case. However numerical simulations with small negative $x_{\text{abs}}$ indicate interesting results.  Figure~\ref{survivalfig2}(a) suggests that the decay of the survival probability at large times  is still exponential with the same decay exponent $\frac{\alpha(\gamma+\alpha)}{2\alpha+\gamma}$ as in \eref{survres3}. The mean first passage time on the other hand shows a non-monotonic behavior with the resetting rate,  Figure~\ref{survivalfig2}(b). Staring from $\alpha=0,$ $T_\alpha$ first decreases, reaches a minimum and then goes up again. This is shown in  Figure~\ref{survivalfig2}(b). This can be explained in the context of a search process as follows. In the absence of resetting the time taken by the RTP to find the target $x=-|x_{\text{abs}}|$ is infinite, however as we increase the resetting rate $\alpha$ to the origin, the RTP comes back to the origin and starts a fresh search. The probability that it finds the target increases and the mean time becomes finite. But if we keep on increasing $\alpha$, then the RTP resets even before it can reach the target, thus the mean first passage time increases. This suggests that if an RTP undergoes resetting and the resetting position and the absorbing boundary are different then there is an optimal resetting rate at which the mean first passage time is minimized.
\section{Other Resetting protocols}\label{ren2}
 In the previous sections we have used a  protocol where simultaneously  the position is reset to the origin and  the orientation is randomized at a constant rate $\alpha$. There can be other protocols, of which an interesting and physically relevant protocol is where along with the position being reset to the origin, the orientation is reset to a fixed direction at a constant rate. 
In that case, due to resetting to a particular orientation (say $\theta_{r}$) the isotropy of the system breaks down and the full distribution depends both on $r,\phi$ where $\phi$ is the polar angle in the $2$d plane. We start with the same initial conditions,\ie, the RTP starts from the origin with a random orientation in $[0,2\pi]$. To investigate this in more detail we try to compute the $x$ and $y$ marginal stationary state distributions starting from a renewal equation for this specific dynamics,
 \bea
 P_{\alpha}(x_i,t)&=&e^{-\alpha t}P_0(x_i,t)+\alpha\int_{0}^{t}ds~e^{-\alpha s}P_0(x_i,s|\theta_{r}),
\label{restheta}
 \eea
where $x_i$ can be $x$ or $y$. Note that, here $P_0(x_i,t)$ denotes the propagator for an RTP without resetting with random initial orientation, while $P_0(x_i,t|\theta_{0})$ is the propagator of an RTP starting with an initial orientation $\theta_0$, \ie, $P_0(x_i,t)=\int_0^{2\pi}d\theta_0 P_0(x_i,t|\theta_{0})$. We have used the same letter for both cases for notational simplicity.

\begin{figure}[h]
\begin{center}
\includegraphics[width=7 cm]{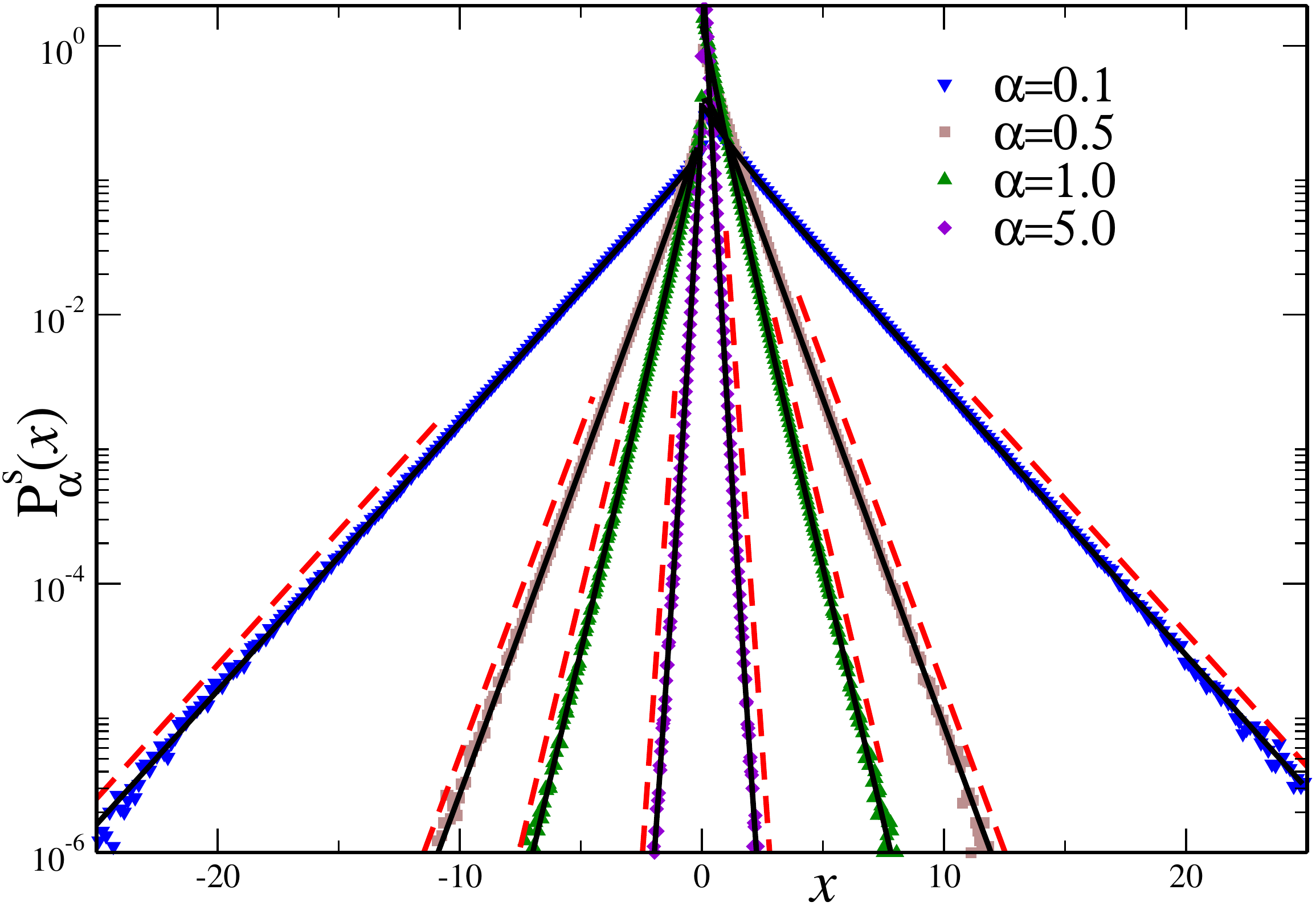}
\includegraphics[width=7 cm]{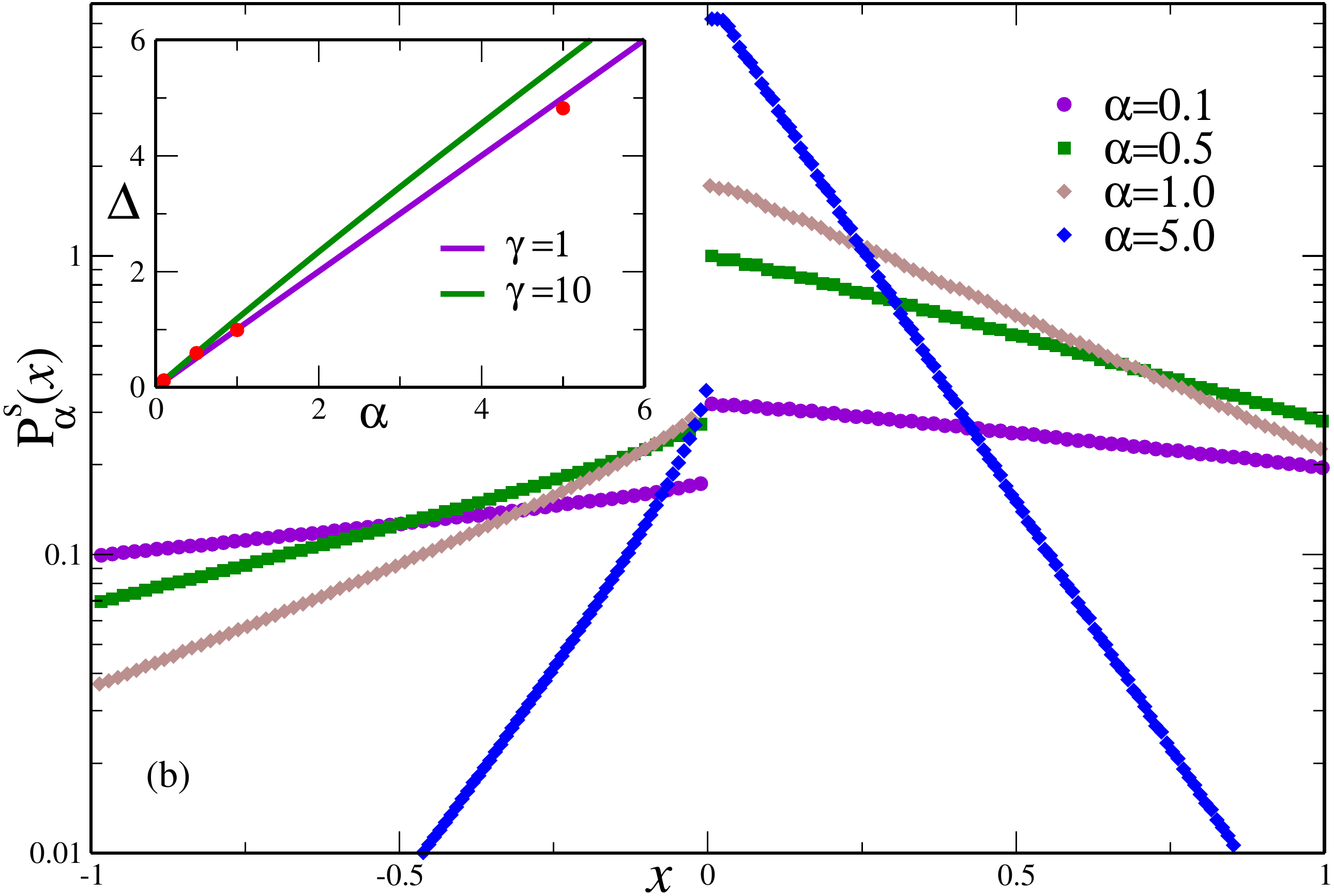}
\caption{Plot of the marginal stationary state distribution for resetting the position and orientation of the RTP to the origin and $\frac{\pi}{4}$ respectively. (a) The colored points are the ones obtained from numerical simulations; the solid black lines represent the analytical form for $P^s(x)$ given in ~\eref{xgreater0}, \eref{xlesser0}; the red dashed lines show the large $x$ behavior of the stationary distribution as given in  \eref{largexthet}. (b) shows a zoomed plot for the same data as in (a) such that the discontinuity across the origin is visible clearly. Solid lines in the inset shows a plot of the discontinuity $\Delta$ against the resetting rate $\alpha$ using \eref{fixedth_disc}; the red points correspond to the value of $\Delta$ calculated from the data in the main Figure~.}
\label{thetresfig}
\end{center}
\end{figure}

The first term in the above equation corresponds to contributions coming from the trajectories with no resetting events, the second term calculates the contributions coming from all the trajectories where the last resetting occurs at time $t-s$ and in the remaining time $s$, there is no more resetting. It is evident from  \eref{restheta} that we need $P_0(x_i,s|\theta_0)$ .  The Fourier-Laplace transform of this propagator, calculated in the \ref{app_fixedth}, is 
\bea
 \tilde{P}_0(k_{i},s|\theta_0)&=&\frac{1}{\gamma+s-ik_iv_0 g_i(\theta_0)}\frac{\sqrt{(s+\gamma)^2+k_i^2v_0^2}}{\sqrt{(s+\gamma)^2+k_i^2v_0^2}-\gamma},
 \label{woresthet}
\eea 
where the subscript $i$ denotes $x,y;$   $g_x(\theta)=\cos\theta$ and $g_y(\theta)=\sin\theta$. 
Since we are interested in the stationary distribution, we take the $t\rightarrow\infty$ limit of  \eref{restheta}, 
\bea
 P_{\alpha}^{s}(x_i)&=&\alpha\int_{0}^{\infty}ds~e^{-\alpha s}P_0(x_i,s|\theta_{r}).
 \label{ssthet1}
 \eea
We can identify the integral on the rhs as a Laplace transform ($s\rightarrow \alpha$) of $P_0(x_i,s|\theta_{r})$. Thereafter taking a Fourier transform w.r.t. $x_i$ on both sides of  \eref{ssthet1} we have
\bea
\hat{P}_{\alpha}^{s}(k_i)&=&\alpha \tilde{P}_{0}(k_i,\alpha|\theta_{r}).
\label{ssthet2}
\eea
Now using  \eref{woresthet} in  \eref{ssthet2} and taking an inverse Fourier transform, we can write the stationary state distributions as
\bea
P_{\alpha}^s(x_i)&=&\frac{\alpha}{2\pi}\int_{-\infty}^{\infty}dk_i\frac{e^{-ik_ix_i}}{\gamma+\alpha-ik_iv_0g_i(\theta_{r})}\frac{\sqrt{(\alpha+\gamma)^2+k_i^2v_0^2}}{\sqrt{(\alpha+\gamma)^2+k_i^2v_0^2}-\gamma}.
\label{largexthetfl}
\eea
This complex integral can be reduced to a simpler form by using the contour integration method in the complex $k_i$-plane and considering the contributions from all the singularities. The details are provided in  \ref{app_disc}, where we obtain an expression for $P_{\alpha}^s(x)$ in terms of a simple real integral (see Equations \eref{xgreater0} and \eref{xlesser0}). The integrals in these equations can be numerically  evaluated for arbitrary values of $x;$  Figure~\ref{thetresfig}(a) compares the prediction with the data obtained from numerical simulations with $\theta_r=\frac \pi 4$, where both the $x$ and $y$ marginal distributions have the same form.

We can also obtain an explicit form for the large-$x$ asymptotic behavior of $P_{\alpha}^s(x).$  At large $x,$ the dominant contribution comes from the poles $k_0=\pm i\sqrt{\alpha^2+2\alpha\gamma},$ for $x_i<0$ and for $x_i>0$ respectively. Computing the residues at these poles, we get, for large $|x_i|,$
\bea
P_{\alpha}(x_i)&\sim &\exp{\left[-\frac{|x_i|}{v_0}\sqrt{\alpha^2+2\alpha\gamma}\right]}
\label{largexthet}
\eea
which is independent of the resetting orientation $\theta_r.$
This is compared with the numerical simulations in  Figure~\ref{thetresfig} (a), in red dashed lines;  a good agreement is observed.  

Note that there is a discontinuous jump in the stationary state distribution across the origin $x=0$. The contribution to the probability distribution near the origin comes for the trajectories which have undergone resetting very close to the observation time. Now, as the RTP is being reset to a particular orientation $\theta_r$ every time (in  Figure~\ref{thetresfig} $\theta_{r}=\pi/4$, which corresponds to a positive velocity and thus taking the RTP away from the origin along positive $x$-axis) the probability that the particle is found in the region $x=0^{-}$ is much smaller compared to the probability for it to be found near $x=0^+$ which results in the discontinuous jump in the probability distribution across the origin.  An explicit form  for the discontinuity $\Delta=P_{\alpha}^{s}(0^{+})-P_{\alpha}^{s}(0^{-})$ near the origin has been obtained in \ref{app_disc} (see \eref{fixedth_disc}). Figure~ \ref{thetresfig}(b)  shows a plot of $\Delta$ as a function of the resetting rate $\alpha$ for a set of values of $\gamma.$
For a fixed $\gamma$ the jump increases with increase in $\alpha$; if $\alpha$ is fixed then the jump increases with increase in $\gamma$, but the increase is very slow, as seen in  Figure~\ref{thetresfig}(b) inset.

\section{Conclusion}\label{concl}

 In this paper we have studied the effect of stochastic resetting on an RTP in two spatial dimensions: the RTP starts from the origin with a random orientation in $[0,2\pi]$ and at a constant rate restarts the process. A stationary state is attained in the long time limit. We compute exactly the radial and $x$-marginal distributions in the stationary state which show a richer behavior than diffusion in the presence of resetting. We  show that both the stationary distributions have exponential tails with the same decay constant. The behavior of the stationary distributions near the origin is governed by the activity where we see a non-vanishing probability density for the radial distribution and a logarithmic divergence for the $x$-marginal distribution.

It turns out that at a finite time, there is a domain in space inside which the NESS has been attained while regions outside it remain in the transient regime. We show that the boundary of this domain propagates linearly with time, so at very large times we expect the distribution to attain stationary-state. The exact analytic expression for the probability distribution have been calculated and they agree with the ones obtained from numerical simulations. 
The presence of stochastic resetting is known to non-trivially change the first passage properties of diffusion processes. In this paper we investigate how the survival probability of the RTP in the presence of an absorbing boundary changes when stochastic resetting is introduced. In particular, we investigate two scenarios, (i) when the particle is reset to a position very close to the absorbing boundary, and (ii) when the resetting position is a finite distance away from the absorbing boundary.  We calculate the survival probability and show that it decays exponentially at large times. 

We investigate the dependence of the mean first passage time on the resetting rate. We show that when the resetting position coincides with the absorbing boundary position, the MFPT monotonically decreases with increasing resetting rate, irrespective of the value of the flipping rate. On the other hand, when the particle is reset to some position away from the absorbing boundary, the MFPT shows a non-monotonic behavior; it reaches a minimum at an optimal resetting rate. 

We also study the RTP dynamics in the presence of resetting to a fixed orientation (along with the position resetting). We compute the stationary position distribution which shows an exponential spatial decay. Moreover, we show that the stationary distribution has a discontinuous jump across the origin, which we also compute exactly. For this fixed orientation resetting protocol, the first passage properties are expected to depend significantly on the resetting orientation. We plan to study this in a future work.

In the spirit of Refs.~\cite{reset_trap,reset_trap2},  the stochastic resetting can be thought of as an effect of an external trap which is switched on and off at random times. It would be interesting to study the effect of such a resetting protocol on RTP dynamics, which also opens up the possibility of experimental realization. Another set of open questions is what happens when an active particle is subjected to non-Markov resetting protocols, \eg, with non-exponential resetting time distributions.

\section{Acknowledgements}
U.B. acknowledges support from Science and Engineering Research Board (SERB), India under Ramanujan Fellowship (Grant No. SB/S2/RJN-077/2018).

\appendix 

\section{Saddle Point Integral}\label{saddlept}
Saddle point integration technique has been used quite extensively in this article. In this Appendix we show the evaluation of $H(r,t)$ (in \eref{radsaddle}) using this method; all the other integrals are also computed in a similar way
\bea
H(r=zv_0t,t)&=&\frac{\alpha\gamma z v_0 t}{v_0^2}\int_{z}^{1} d\tau~e^{-t\phi(z,\tau) }\frac{1}{\sqrt{\tau^2-z^2}} \label{eq:Hrt}
\label{saddleapp1}
\eea
 with  $\phi(z,\tau)=(\alpha+\gamma)\tau-\gamma\sqrt{\tau^2-z^2}$. The function $\phi(z,\tau)$ always has a minimum w.r.t. $\tau$ at some $\tau_0$,  which is obtained by solving 
\bea
\left.\frac{\partial \phi(z,\tau)}{\partial\tau}\right|_{\tau=\tau_0}=0 \Rightarrow \tau_0 = \frac{z(\alpha+\gamma)}{\sqrt{\alpha^2+2\alpha\gamma}}.
\eea
Since the denominator of the integrand in  \eref{eq:Hrt} is  a monotonically decreasing function for $\tau>z$, the integrand has a maximum at $\tau=\tau_0$. For large $t$, the integrand becomes sharply peaked at $\tau_0$, so we can expand $\phi(z,\tau)$ in a Taylor series about  $\tau=\tau_0$,
\bea
\phi(z,\tau)=\phi(z,\tau_0)+\frac{(\tau-\tau_0)^2}{2} \phi''(z,\tau_0)+\cal O \left[(\tau-\tau_0)^3\right],
\eea
where $^\prime$ denotes derivative w.r.t. $\tau.$  Keeping upto the quadratic term in the expansion for $\phi(z,\tau)$, gives a very good estimate of the integral in \eref{saddleapp1} at large $t$.
 
 \begin{figure}[h]
 \includegraphics[width=8 cm]{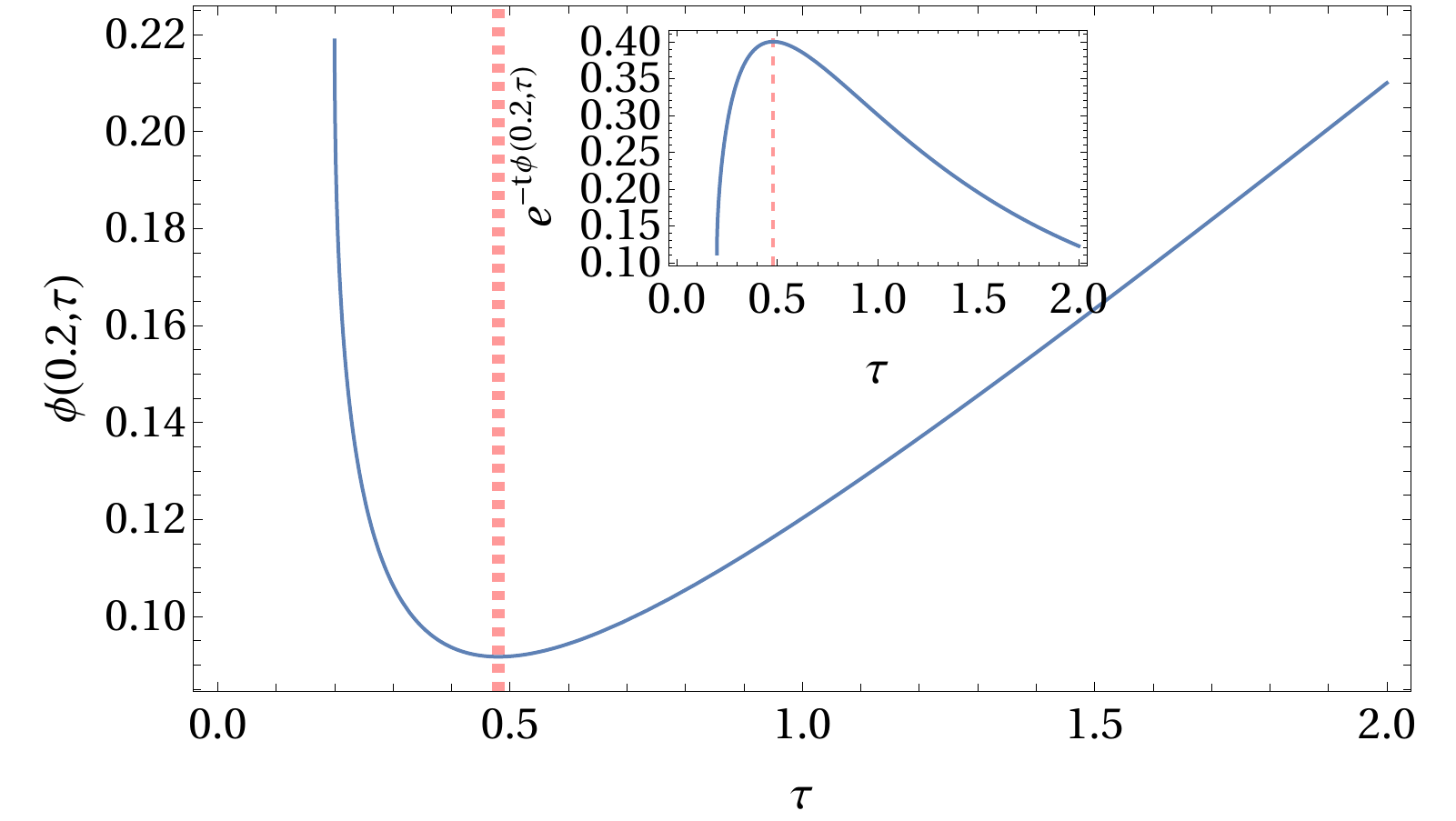}
 \includegraphics[width=8 cm]{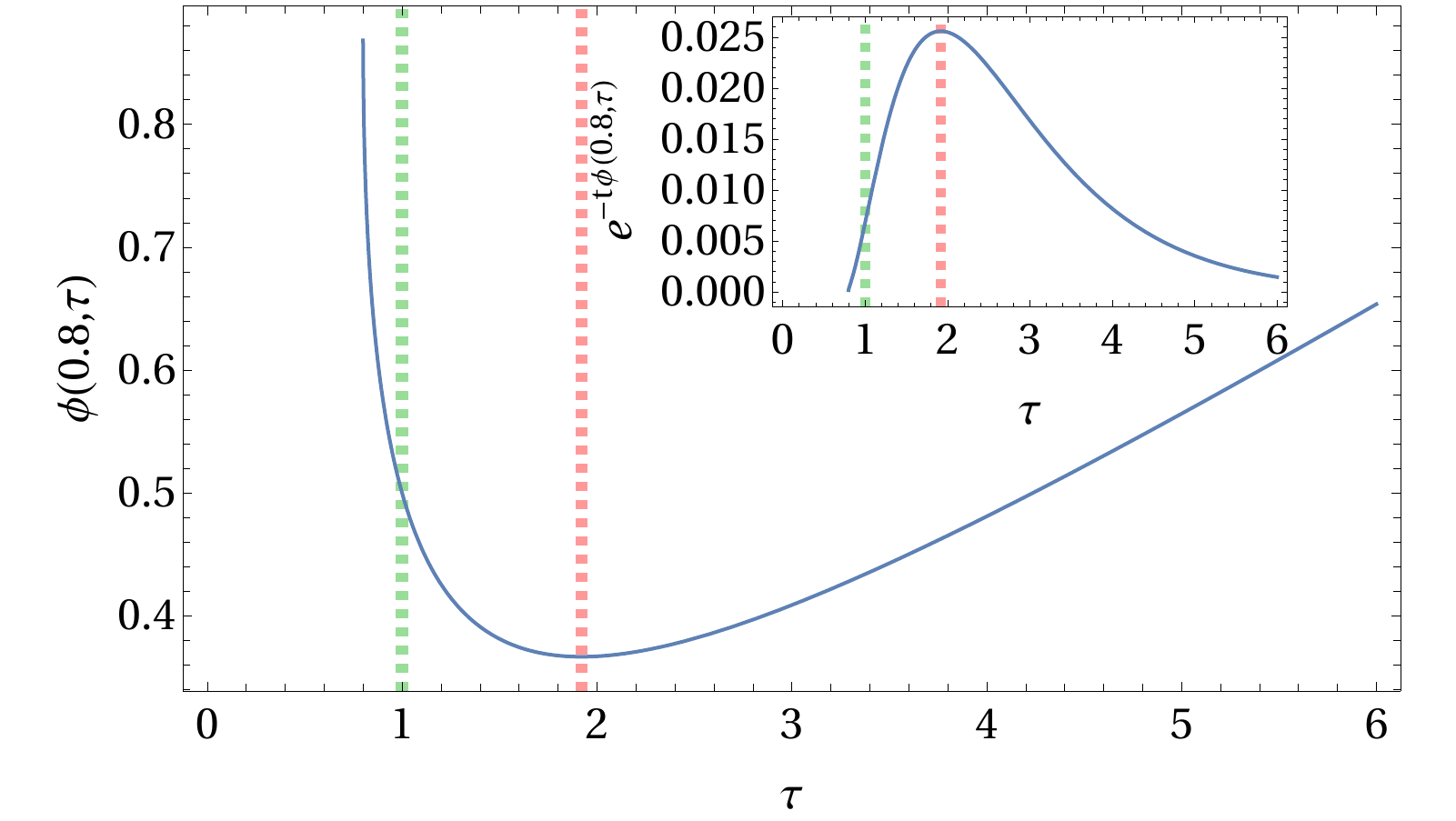}
 \caption{Plot of $\phi(z,\tau)$ and $\e^{-t\phi(z,\tau)}$ for $\alpha=0.1;\gamma=1$: Left panel shows a case where  $z<\frac{(\alpha+\gamma)}{\sqrt{\alpha^2+2\alpha\gamma}}$, while the  right panel shows a case where $z>\frac{\sqrt{\alpha^2+2\alpha\gamma}}{(\alpha+\gamma)}$. The red dashed lines denote the actual minimum of the functions while the green dashed ones on the right panel denote the point $\tau=1$. We have taken $t=10$  in the insets. The peak becomes sharper as we keep increasing $t$.}
 \label{saddleappendixfig}
 \end{figure}
 
  This form of $\phi(z,\tau)$ is used when $z<\tau_0<1$, \ie, $\tau_0$ lies within the integration limits \ie, $[z,1]$.
  \bea
\fl H(r=zv_0t,t)\approx \frac{\alpha \gamma z v_0 t}{\sqrt{2}v_0^2}\frac{e^{-t\phi(z,\tau_0)}}{\sqrt{\tau_0 ^2-z^2}}\int_{-\infty}^{\infty}d\tau'~ \e^{-t\phi''(z,\tau_0)\tau'^2}=\frac{\alpha \gamma\sqrt{zv_0t} \sqrt{\pi/2}}{v_0^2(\alpha^2+2\alpha\gamma)^{1/4}}\e^{-zt\sqrt{\alpha^2+2\alpha\gamma}}\cr
  \eea
   If $\tau_0>1$ ie, $z>\frac{\sqrt{\alpha^2+2\alpha\gamma}}{(\alpha+\gamma)}$ , then within the integration limits the integrand reaches its maximum value at the boundary $\tau_0=1$ as shown in  \ref{saddleappendixfig}. In such a scenario we can still expand $\phi(z,\tau)$ about $\tau=1$ as,
 \bea
  \phi(z,\tau)=\phi(z,1)+(\tau-1) \phi'(z,1)+ \cal O \left[(1-\tau)^2\right]
 \eea
At large $t$ we keep upto the second term in the above expansion and have
 \bea
 \fl H(r=zv_0t,t)&\approx&\frac{\alpha \gamma z v_0 t}{\sqrt{2}v_0^2}\frac{e^{-t\phi(z,1)}}{\sqrt{1-z^2}}\int_{z}^{1}d\tau \e^{-t(1-\tau)|\phi'(z,1)|}=A(z) e^{-(\alpha+\gamma)t+\gamma t\sqrt{1-z^2}}
 \label{hrtsaddle}
 \eea
 where 
 \bea
\fl A(z)=\frac{\alpha \gamma z }{\sqrt{2}v_0(1-z^2)^{1/2}}\frac{\left(1-\e^{-t(1-z)|\phi '(z,1)|}\right)}{|\phi'(z,1)|}\cr \fl\qquad =\frac{\alpha \gamma z}{\sqrt{2}v_0  \left(\gamma-(\alpha+\gamma)\sqrt{1-z^2}\right)}\left(1-\exp\left[-t\sqrt{\frac{1-z}{1+z}}\left(\gamma-(\alpha+\gamma)\sqrt{1-z^2}\right)\right]\right).\cr
\label{a7}
\eea 
We used $\phi '(z,1)=-\left[\frac{\gamma-(\alpha+\gamma)\sqrt{1-z^2}}{\sqrt{1-z^2}}\right]$ to arrive at the final line. Note that for $z>z^*$, $\left(\gamma-(\alpha+\gamma)\sqrt{1-z^2}\right)>0$.
 
 The exponential part in \eref{hrtsaddle} gives the leading order contribution to $H(r,t)$, while $A(z,t)$ gives a sub-leading contribution. This along with the 2nd term on the rhs of \eref{renrfull} is plotted in Figure \ref{nessclip} (b) with magenta lines.
 
\section{Free propagator for RTP starting with fixed orientation}\label{app_fixedth}

In this section we derive the Fourier Laplace transformation of the $x$-marginal distribution of the 2$d$ RTP starting from the origin with a fixed orientation $\theta_0$.
\begin{figure}[h]
\centering\includegraphics[width=8 cm]{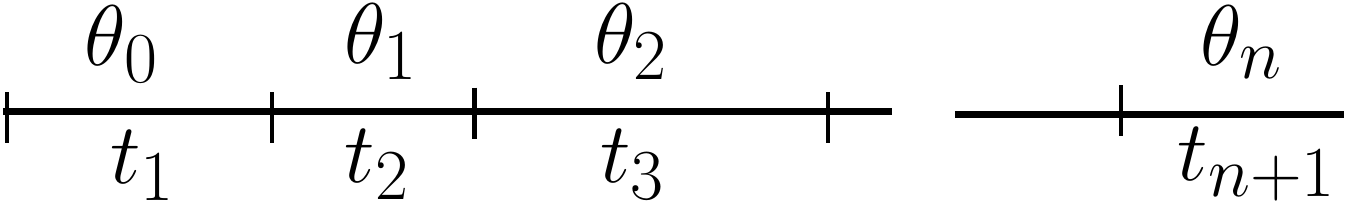}
\caption{Schematic representation of a typical trajectory with $n$ flips; the total duration $t$ is divided into $n+1$ intervals $\{t_i; i=1,2,\cdots n+1 \}$; $\theta_{i-1}$ denotes the orientation in the $i^{\text{th}}$ interval.}
\label{interval}
\end{figure}

 We consider a trajectory with $n$ orientation flips during the time interval $[0,t]$. Let $t_i$ denote the time-interval between $(i-1)^{\text{th}}$ and $i^{\text{th}}$ flips, then $t= \sum_{i=1}^{n+1} t_i$ where $t_{n+1}$ is the time between the last flip and the final time. Also, let  $\theta_{i-1}$ denote the orientation during the $i^{th}$ interval (see Figure~\ref{interval} for a schematic representation). Then, the final position $x= v_0 \sum_{i=1}^{n+1} t_i \cos \theta_{i-1}.$ The position distribution is obtained by considering all such possible trajectories. It is convenient to consider the generating function of the position distribution,
 \bea
\fl \qquad \langle \e^{ik_x x}\rangle=\sum_{n=0}^{\infty}\gamma^n \e^{-\gamma t}\int_{0}^{\infty}\prod_{i=1}^{n+1} dt_i \int_0^{2\pi} \prod_{i=1}^{n}d\theta_i ~ \exp{\left[i v_0 k_x \sum_{i=0}^{n}\cos\theta_i\right ]}~\delta\left(t-\sum_{i=1}^{n+1} t_i\right). \n
\eea
Taking a Laplace transform of the above equation w.r.t. time, we have,
\bea
\fl \tilde{P}_0(k_x,s|\theta_0) = \int_0^\infty dt~ e^{-s t}  \langle \e^{ik_x x}\rangle\cr
\fl \qquad \qquad \quad =\sum_{n=0}^{\infty}\gamma^n \int_{0}^{\infty}dt_1~\e^{-(\gamma+s-i v_0 k_x \cos\theta_0)t_1}\left( \int_{0}^{\infty}dt\int_{0}^{2\pi}d\theta~ \e^{-(\gamma +s-i v_0 k_x \cos\theta)t}\right)^n \n
\eea
The $\theta$ and $t$ integral can both be computed exactly, and yields,
\bea
\fl \tilde{P}_0(k_x,s|\theta_0) =\frac{1}{\gamma+s-ik_xv_0\cos\theta_0}\sum_{n=0}^{\infty}\left(\frac{\gamma}{\sqrt{(s+\gamma)^2+v_0^2k_x^2}}\right)^n\cr
=\frac{1}{\gamma+s-ik_xv_0\cos\theta_0}\frac{\sqrt{(s+\gamma)^2+v_0^2k_x^2}}{\sqrt{(s+\gamma)^2+v_0^2k_x^2}-\gamma}.
\eea
This is the result used in the main text.

\section{Stationary State for resetting to fixed orientation $\theta_r$}\label{app_disc}
Let us  rewrite \eref{largexthetfl} in the main text for the $x$-component only, 
\bea
P_{\alpha}^s(x)&=&\frac{\alpha}{2\pi}\int_{-\infty}^{\infty}dk_x\frac{e^{-ik_x x}}{\gamma+\alpha-ik_x v_0\cos\theta_{r}}\frac{\sqrt{(\alpha+\gamma)^2+k_x^2v_0^2}}{\sqrt{(\alpha+\gamma)^2+k_x^2v_0^2}-\gamma}.
\label{app_fl}
\eea
The integral is along the real line, so we can use semicircular contours to evaluate the integral. The integrand for $P^{s}_{\alpha}(x)$ has three poles at $\pm i\sqrt{\alpha^2+2\alpha\gamma}$ and $-i(\gamma+\alpha)\sec\theta_r$; and two branch-points at $\pm i(\alpha+\gamma).$ 

For $x>0$, the contour has to be closed on the lower half plane to make the arc integral vanish for large $|k_x|$ in accordance with Jordan's lemma. The line integral \eref{app_fl} has contributions from the poles at $- i\sqrt{\alpha^2+2\alpha\gamma}$, $-i(\gamma+\alpha)\sec\theta_r$ and the branch cut along the imaginary $k_x$-axis from $-i(\alpha+\gamma)$ to $-\infty$. Calculating all these contributions and using Cauchy's residue theorem we have, for $x>0,$ 
\bea
\fl P^s_{\alpha}(x)=\frac{\alpha(\alpha+\gamma)^2\e^{-\frac{x}{v_0}(\alpha+\gamma)\sec\theta_r}\tan ^2\theta_r}{v_0\left(\gamma^2+(\alpha+\gamma)^2\tan^2\theta_r\right)}+\frac{\alpha\gamma^2}{v_0}\frac{e^{-\frac{x}{v_0}\sqrt{\alpha^2+2\alpha\gamma}}}{\sqrt{\alpha^2+2\alpha\gamma}(\alpha+\gamma-\sqrt{\alpha^2+2\alpha\gamma}\cos\theta_r)}\cr+\frac{\alpha\gamma}{\pi v_0}\int_0^{\infty}\frac{\e^{-\frac{|x|}{v_0}(\alpha+\gamma+u)}}{(\alpha+\gamma)(1-\cos\theta)-u\cos\theta}\frac{\sqrt{u^2+2u(\alpha+\gamma)}}{\left(u^2+2u(\alpha+\gamma)+\gamma^2\right)}.
\label{xgreater0}
\eea
Similarly, for $x<0$ the contour has to be closed on the upper half plane to make the arc integral vanish.  The line integral in \eref{app_fl} in this case has contributions from the pole at $i\sqrt{\alpha^2+2\alpha\gamma}$ and the branch cut along the imaginary $k_x$-axis from $i(\alpha+\gamma)$ to $\infty$. Calculating all these contributions and using Cauchy's residue theorem we have, for $x<0,$
\bea
\fl P^s_{\alpha}(x)=\frac{\alpha\gamma^2}{v_0}\frac{e^{-\frac{|x|}{v_0}\sqrt{\alpha^2+2\alpha\gamma}}}{\sqrt{\alpha^2+2\alpha\gamma}(\alpha+\gamma-\sqrt{\alpha^2+2\alpha\gamma}\cos\theta_r)}\cr+\frac{\alpha\gamma}{\pi v_0}\int_0^{\infty}du~\frac{\e^{-\frac{|x|}{v_0}(\alpha+\gamma+u)}}{(\alpha+\gamma)(1+\cos\theta)+u\cos\theta}~\frac{\sqrt{u^2+2u(\alpha+\gamma)}}{\left(u^2+2u(\alpha+\gamma)+\gamma^2\right)}.
\label{xlesser0}
\eea 
Though the expressions \eref{xgreater0} and \eref{xlesser0} are fairly complicated it can be very easily seen that $P^s_{\alpha}(x)$ has a discontinuity at $x=0.$ The discontinuity $\Delta=P_{\alpha}^{s}(0^{+})-P_{\alpha}^{s}(0^{-})$ is given by
\bea
\fl\Delta&=\frac{\alpha(\alpha+\gamma)^2\tan ^2\theta_r}{v_0\left(\gamma^2+(\alpha+\gamma)^2\tan^2\theta_r\right)}
\cr\fl &-\frac{2\alpha\gamma\cos{\theta_{r}}}{\pi v_0}\int_{0}^{\infty} du\frac{\sqrt{u^2+2u(\alpha+\gamma)}(\alpha+\gamma+u)}{\left(u^2+2u(\alpha+\gamma)+\gamma^2\right)\left((\gamma+\alpha)^2-\cos^2\theta_{r}(\gamma+\alpha+u)^2\right)}
\label{fixedth_disc}
\eea
The integrals in \eref{xgreater0},~\eref{xlesser0} and \eref{fixedth_disc} can be evaluated numerically and have been shown in the Figure~\ref{thetresfig} in the main text.

\end{document}